\documentclass[aps,pra,preprint,unsortedaddress,superscriptaddress]{revtex4-2}
\usepackage{times}
\usepackage{graphicx,float} 
\usepackage{dcolumn}
\usepackage{bm}
\usepackage{amsmath,amssymb,amsfonts}
\usepackage{amsthm}
\usepackage{mathrsfs}
\usepackage{makecell}
\usepackage{multirow}
\usepackage{booktabs}
\usepackage{tikz}
\usetikzlibrary{decorations.pathmorphing}
\usetikzlibrary{arrows.meta}
\usepackage{listings}
\usepackage{lipsum} 
\usepackage{xcolor}
\usepackage{textcomp}

\theoremstyle{plain}
\newtheorem{theorem}{Theorem}
\newtheorem{lemma}[theorem]{Lemma}

\theoremstyle{definition}
\usepackage[colorlinks,linkcolor=blue,anchorcolor=blue,citecolor=blue,urlcolor=red]{hyperref}

\newcommand{\casql}{Laboratory of Quantum Information, University of Science and Technology of China, Hefei, Anhui, 230026, China}
\newcommand{\casex}{Anhui Province Key Laboratory of Quantum Network, University of Science and Technology of China, Hefei 230026, China}

\newcommand{\aihf}{Institute of Artificial Intelligence, Hefei Comprehensive National Science Center, Hefei, Anhui, 230088, China}
\newcommand{\origin}{Origin Quantum Computing Technology (Hefei) Co., Ltd., Hefei, Anhui, 230026, China}

\newcommand{\norm}[1]{\left\|#1\right\|}

\newcommand{\Ai}{\operatorname{Ai}}
\newcommand{\ii}{\mathrm{i}}
\newcommand{\e}{\mathrm{e}}
\newcommand{\eps}{\epsilon}
\hypersetup{pdftitle={Quantum Simulation of Non-Hermitian Special Functions and Dynamics via Contour-based Matrix Decomposition},pdfauthor={Chao Wang et al.}}
\begin{document}
	\title{Quantum Simulation of Non-Hermitian Special Functions and Dynamics via Contour-based Matrix Decomposition}
	\author{Chao Wang}
	\affiliation{\origin}
	\author{Huan-Yu Liu}
	\email{liuhuanyu@ustc.edu.cn}
	\affiliation{\aihf}
	\author{Cheng Xue}
	\affiliation{\aihf}
	\author{Xi-Ning Zhuang}
	\affiliation{\origin}
	\affiliation{\casql}
	\affiliation{\casex}
	\author{Menghan Dou}
	\affiliation{\origin}
	\author{Zhao-Yun Chen}
    \email{chenzhaoyun@iai.ustc.edu.cn}
	\affiliation{\aihf}
	\author{Guo-Ping Guo}
	\affiliation{\origin}
	\affiliation{\aihf}
	\affiliation{\casql}
	\affiliation{\casex}
	\date{Revised draft: November 5, 2026}

	\begin{abstract}
Simulating non-Hermitian dynamics on quantum computers is often hindered by the decay of success probability and the treatment of non-diagonalizable matrices. Here, we present contour-based matrix decomposition (CBMD), a quantum functional calculus framework for evaluating suitable non-Hermitian matrix functions. By applying the matrix-valued Cauchy residue theorem, CBMD expresses the target function as a contour-residue identity, followed by finite truncation with controlled error to yield linear combinations of functions of Hermitian matrices. For dissipative first-order dynamics, we query the original block-encoding of $A(t)/\alpha_A$. The construction has constant coefficient weight and a logarithmic sampling cutoff, matching the leading precision dependence of optimal LCHS kernels in the same comparison. For the second-order wave, Bessel $J_0$, and Airy equations with $A=L+iH$ and $L,H\succ0$, the matrix inputs are two separate block-encodings, one of $H/\norm H$ and one of $L/\norm L$. The resulting normalization envelope is governed by the anti-Hermitian part and can improve substantially over a direct Taylor--LCU construction. The complete query bounds also retain spectral conditioning, output-state amplification, and algebraic precision dependence from residue truncation. CBMD avoids diagonalizing the non-Hermitian matrix. Its efficiency requires suitable analytic growth bounds and efficient input access.
\end{abstract}
\maketitle
\tableofcontents
\clearpage
\section{Introduction}\label{sec1}
	
Quantum algorithms have been developed for solving linear systems~\cite{PhysRevLett.103.150502, doi:10.1137/16M1087072}, unstructured search~\cite{10.1145/237814.237866}, and the simulation of quantum many-body systems~\cite{RevModPhys.86.153, Altman2021qsim, daley2022practical}. While evaluating functions of Hermitian matrices and simulating unitary dynamics represent a natural fit for quantum hardware, many frontier physical and mathematical problems require evaluating complex holomorphic functions of non-Hermitian operators. Non-Hermitian matrix functions describe effective dissipative dynamics and give analytical solutions to many matrix differential equations~\cite{an2024laplacetransformbasedquantum, PhysRevLett.131.150603,jin2025schrodingerizationmethodlinearnonunitary,jin2026transmutationbasedquantumsimulation, an2023quantumalgorithmlinearnonunitary, yang2025quantumdifferentialequationsolvers, JIN2022111641, PhysRevA.108.032603, childs2020quantum, Berry_2014, berry2017quantum, Berry2024quantumalgorithm, PhysRevApplied.19.054012}.

Quantum singular value transformation (QSVT) implements bounded polynomial transformations of singular values~\cite{10.1145/3313276.3316366}. Matrix polynomials and matrix differential equations also require functions of the matrix itself~\cite{doi:10.1137/1.9780898717778}. For diagonalizable matrices these functions act on eigenvalues. For Jordan blocks they also involve derivatives of the scalar function. Quantum eigenvalue processing applies to nonnormal matrices, including matrices with Jordan blocks. In the diagonalizable case, some of its complexity bounds involve the condition number of the eigenvector basis~\cite{10756112}. Classical matrix functional calculus and contour quadrature express matrix functions in terms of resolvents~\cite{680b6387-6345-3b49-9027-7805b1e240ba,doi:10.1137/070700607}. Quantum contour algorithms implement these resolvents by matrix inversion, with costs depending on resolvent bounds and the input model~\cite{Takahira_2020,takahira2021quantumalgorithmsbasedblockencoding,jiang2026contourintegralbasedquantumeigenvalue}. Direct monomial linear combinations provide another construction, but their coefficient $L_1$-norms can grow exponentially~\cite{BERRY_CHILDS_CLEVE_KOTHARI_SOMMA_2017,10756112,2012}.

For the exponential $e^{-At}$, Schr\"{o}dingerization~\cite{PhysRevA.108.032603,jin2025schrodingerizationmethodlinearnonunitary,PhysRevLett.133.230602} and linear combination of Hamiltonian simulation (LCHS)~\cite{PhysRevLett.131.150603,an2023quantumalgorithmlinearnonunitary,low2025optimalquantumsimulationlinear} address non-unitary evolution under their respective access and stability assumptions. For suitable dissipative problems these methods give query bounds independent of the condition number of the eigenvector basis of $A$. Inspired by LCHS, we use QSVT and linear combinations of unitaries (LCU) to treat the special functions studied below.

In this work, we propose the CBMD method for suitable holomorphic functions whose growth is controlled on the complex contours. By applying the matrix-valued residue theorem, CBMD decomposes a non-Hermitian matrix function into contour and residue terms~\cite{Stein2003}. An exact infinite series follows once the contour remainder vanishes and the residue sum converges. Each retained Hermitian matrix function can be implemented with QSVT when a globally bounded polynomial approximation is available. This approach avoids an explicit eigenvector decomposition.

For wave and Bessel dynamics the target functions are bounded on the positive real sampling spectra. At fixed $t>0$ and bounded $\norm H$, the direct Taylor coefficient sum grows on the exponential scale $e^{t\sqrt{\norm L+\norm H}}$ as $\norm L$ increases. The corresponding CBMD normalization is bounded by $e^{t\sqrt{\norm H/2}}$ times a logarithmic coefficient factor. The complete matrix-query bounds also include polynomial approximation, residue truncation and output-state amplification.

For first-order non-unitary dynamics, the coefficient weight and real cutoff match the leading precision scaling of optimal LCHS kernels~\cite{PhysRevLett.131.150603,an2023quantumalgorithmlinearnonunitary,low2025optimalquantumsimulationlinear}. Other algorithms may improve different resources under different assumptions. Beyond exponentials, we derive a common residue construction for wave, Bessel and Airy functions. Their complexity is stated separately from the coefficient normalization.

By replacing the contour representation with a discrete residue series, CBMD connects classical complex analysis to Hermitian matrix transformations on quantum computers. The Airy application additionally uses a $\mathbb Z_3$ root-of-unity projection to encode the required initial conditions without choosing a matrix cube-root branch.

The framework inherits the costs of input block-encodings, QSVT and LCU, including output-state amplification. The examples below establish sufficient analytic conditions for each of the specified functions. A comparison of total query costs requires all terms in the corresponding theorems.

\section{Contour-based matrix decomposition}\label{sec3}
The spectral theorem reduces a function of a Hermitian matrix to its action on the real eigenvalues. In a quantum circuit this action can be approximated by QSVT when a suitable bounded polynomial is available~\cite{10.1145/3313276.3316366}. CBMD uses complex analysis to express a function of $A=L+iH$ in terms of functions of Hermitian matrices.

For a matrix-valued meromorphic function $F(z)$, apply the scalar residue theorem entry by entry~\cite{Stein2003}. This gives the following identity.
{ 
    \begin{lemma}[Matrix Residue Theorem]
		\label{matrix_rcontour}
		 Let $\Gamma$ be a positively oriented, piecewise $C^1$ simple closed contour. Suppose the entries of $F$ are meromorphic on an open set containing $\Gamma$ and its interior and have no pole on $\Gamma$. If $\{z_k\}$ lists all enclosed poles, then
		\begin{equation}
			\oint_\Gamma F(z) \,\mathrm{d}z = 2\pi i  \sum_{k} \mathrm{Res}(F, z_k),
		\end{equation}
		where $\mathrm{Res}(F,z_k)$ is the matrix coefficient of $(z-z_k)^{-1}$ in the local Laurent expansion.
	\end{lemma}}
	Let $A=L+iH$, with $L,H$ Hermitian. Suppose $f$ is holomorphic on an open set containing the spectra of $L+zH$ for every $z$ on and inside $\Gamma$. Near each fixed $z$, the Dunford integral over a fixed contour in this open set around the spectrum shows that $z\mapsto f(L+zH)$ is holomorphic~\cite{doi:10.1137/1.9780898717778}. We introduce holomorphic functions $h_1(z)$ with real zeros $\{q_r\}$ and an auxiliary holomorphic function $h_2(z)$ with non-real zeros $\{p_r\}$. All denominator zeros are assumed simple and mutually distinct, $h_1(i)h_2(i)\ne0$, the target pole $i$ lies inside $\Gamma$, and no denominator zero lies on $\Gamma$.
	
	Consider the matrix-valued function
	\begin{equation}
		F(z) = \frac{h_{1}( i )h_{2}( i )f({zH+L})}{(z- i )h_1(z)h_2(z)},
	\end{equation}
	whose possible poles belong to $S = \{ i \}\cup\{p_r\}\cup\{q_r\}$. All residue sums below run over the indicated denominator zeros inside $\Gamma$, with zero residue at any removable singularity. Applying {Lemma}~\ref{matrix_rcontour} gives
	{\begin{equation}
	\underbrace{\frac{1}{2\pi  i }\oint_\Gamma F(z) \,dz}_{\text{Contour remainder}}
			=\underbrace{f( i H+L)}_{\text{Target Transformation}} + \underbrace{\sum_{s\in\{q_r\}}{c_sf(H_s)}}_{\text{Discrete Sampling (QSVT+LCU)}} + \underbrace{\sum_{s\in \{p_r\}}{c_sf(H_s)}}_{\text{Auxiliary residues}},  \label{eq:lcf}
	\end{equation}}
	where {$H_s=sH+L$} and
	$c_s=\lim\limits_{z\rightarrow s}\frac{(z-s)h_{1}( i )h_{2}( i )}{(z- i )h_1(z)h_2(z)}$. For a finite contour this is an exact finite-residue identity. An infinite-series representation additionally requires a specified sequence of contours along which the remainder tends to zero and the residue series converges. The error quantities are
\begin{equation}
\epsilon_{\rm IR}=\left\|\frac1{2\pi i}\oint_\Gamma F(z)\,dz\right\|,
\qquad \epsilon_{\rm AE}=\left\|\sum_{s\in\{p_r\}}c_sf(H_s)\right\|.
\end{equation}
Solving Eq.~\eqref{eq:lcf} for the target places minus signs in front of both residue sums. When the two error terms are small, $f(A)$ is approximated by $-\sum_s c_sf(H_s)$ over real sampling poles. Alternative substitutions, such as $f(iH+izL)$ with target pole $-i$, use the same principle.

Because $H_s=sH+L$ is Hermitian for real $s$, each retained term can be evaluated by Hamiltonian simulation or QSVT under the corresponding polynomial approximation conditions~\cite{10.1145/3313276.3316366,PhysRevLett.118.010501,Low2019hamiltonian}. LCU combines these terms~\cite{BERRY_CHILDS_CLEVE_KOTHARI_SOMMA_2017}. Its normalization must include both the coefficient magnitudes and the normalization of each implemented function. Thus bounded modulus on the sampling spectra is useful, but does not by itself bound the polynomial degree.

For example, $|e^{iz}|=1$ for real $z$, so sampling the exponential of a Hermitian matrix gives a unitary~\cite{an2023quantumalgorithmlinearnonunitary,low2025optimalquantumsimulationlinear}. Applications II and III instead use a bound on the positive real axis together with a positive spectral gap.

\subsection{Input access and the two LCU layers}
The contour identity above is independent of the matrix input model. For first-order non-unitary dynamics (Application I), the sole matrix oracle is the original block-encoding $U_A$ of the generator, with inverse and controlled access. Suppressing its time argument, we write
\begin{equation}\label{eq:input}
(\langle0^{a_0}|\otimes I)U_A(|0^{a_0}\rangle\otimes I)=A/\alpha_A,\qquad \alpha_A>0,\quad \alpha_A\ge\norm A.
\end{equation}
For $A(t)$, the same oracle is queried coherently at the required times. Starting with the wave equation, the wave, Bessel $J_0$, and Airy algorithms instead take $H/\norm H$ and $L/\norm L$ as their two primitive matrix blocks [Eq.~\eqref{eq:separate-oracles}]. In both access models the query statements exclude classical coefficient computation, reversible arithmetic, and state-preparation gate counts. For $u_0\ne0$, the normalized input $|u_0\rangle=u_0/\norm{u_0}$ is provided by an oracle and its inverse. The input includes a known positive lower bound on the norm of the output for this \emph{normalized} input, i.e., on $\norm{u(t)}/\norm{u_0}$ for a propagator. For bounds expressed using the true ratio, we assume it is at most a known universal constant times the supplied lower bound. If the target vector is zero, a normalized output state is undefined. State-preparation guarantees are conditional on a heralded success outcome; vector-norm error is measured after fixing the physically irrelevant global phase.

To illustrate the single-oracle construction, consider the real sample $M_s=H_s=L+sH$. Its decomposition using $U_A$ is
\begin{equation}\label{eq:Ms}
M_s=\frac{1-\ii s}{2}A+\frac{1+\ii s}{2}A^\dagger,
\qquad \alpha_s=\alpha_A\sqrt{1+s^2}.
\end{equation}
Let $P_0|0\rangle=(|0\rangle+|1\rangle)/\sqrt2$, $\phi=\arctan s$, and
\begin{equation}
P_s|0\rangle=\frac{\e^{-\ii\phi}|0\rangle+\e^{\ii\phi}|1\rangle}{\sqrt2},\quad
S_{\rm in}=|0\rangle\langle0|\otimes U_A+|1\rangle\langle1|\otimes U_A^\dagger.
\end{equation}
Then $(P_0^\dagger\otimes I)S_{\rm in}(P_s\otimes I)$ block-encodes $M_s/\alpha_s$. Application I uses the same two-branch circuit for $H(t)+sL(t)$, with the coefficients in Eq.~\eqref{eq:first-generator}. For the second-order equations the corresponding sample block uses the two primitive inputs directly via Eq.~\eqref{eq:separate-oracles}.

The outer LCU is common to both access models. Write $[V_{s_k}]_{00}$ for the block obtained with all its ancillas in zero, padding the workspaces to a common size. Suppose $\|b_k[V_{s_k}]_{00}-f(M_{s_k})\|\le\delta_k$ with $b_k>0$. To implement a finite sum $\sum_k w_k f(M_{s_k})$, omit zero weights and, for a nonempty sum, set
\begin{equation}\label{eq:outerLCU}
\alpha_{\rm out}=\sum_k|w_k|b_k,\quad
P_{\rm out}|0\rangle=\sum_k\sqrt{\frac{|w_k|b_k}{\alpha_{\rm out}}}\e^{\ii\arg w_k}|k\rangle,
\quad P_{{\rm out},0}|0\rangle=\sum_k\sqrt{\frac{|w_k|b_k}{\alpha_{\rm out}}}|k\rangle.
\end{equation}
Select $V_{s_k}$ with $S_{\rm out}=\sum_k|k\rangle\langle k|\otimes V_{s_k}$. The circuit
$(P_{{\rm out},0}^\dagger\otimes I)S_{\rm out}(P_{\rm out}\otimes I)$ has top block
\[
\frac{1}{\alpha_{\rm out}}\sum_k w_k b_k[V_{s_k}]_{00}.
\]
 Multiplying by $\alpha_{\rm out}$ gives an approximation to the retained sum with error at most $\sum_k|w_k|\delta_k$. Its success probability on $|u_0\rangle$ is the squared norm of the resulting unnormalized output vector divided by $\alpha_{\rm out}^2$. Suppose the contour remainder, auxiliary residues, discarded real samples, and implementation errors together contribute at most $\norm{f(A)|u_0\rangle}/2$ on that input. The triangle inequality then gives success amplitude at least $\norm{f(A)|u_0\rangle}/(2\alpha_{\rm out})$. With the stated norm estimate, amplitude amplification uses $O(\alpha_{\rm out}/\norm{f(A)|u_0\rangle})$ calls for constant heralded success probability~\cite{Brassard2002Amplitude}.

QSVT additionally requires a polynomial bounded on the full interval $[-1,1]$ and appropriate parity handling. Boundedness only on the spectrum of $M_s$ is insufficient. Appendix~\ref{proofthm5} supplies the approximation argument used for the three special-function applications. Their fixed QSVT normalization is included explicitly.

\section{Application I: First-Order Non-Unitary Dynamics}\label{sec1simulation}

	Consider the dynamics of a vector $u(t)$ governed by a linear ordinary differential equation (ODE):
	\begin{equation}\label{eqn:ODE}
		\frac{\mathrm{d} u(t)}{\mathrm{d} t} = -A(t) u(t) + b(t), \quad u(0) = u_0.
	\end{equation}
	Inhomogeneous equations with $b(t)\not\equiv 0$ can be mapped to a homogeneous form~\cite{PhysRevA.108.032603, jin2025schrodingerizationmethodlinearnonunitary, JIN2022111641}, but the embedding can change dissipativity and input costs. The bounds here therefore cover only $b(t)\equiv0$, with $A(t)=L(t)+iH(t)$, $L(t)\succeq0$ pointwise and bounded piecewise-continuous coefficients. This case yields the solution
	\begin{equation}\label{eqn:s_hODEs}
		u(t) = \mathcal{T}\mathrm{e}^{-\int_0^t A(s) \mathrm{d} s} u_0, 
	\end{equation}
	where $\mathcal{T}$ is the time ordering operator. It has wide applications, including simulating imaginary time evolution~\cite{motta2020determining,mcardle2019variational}, open quantum systems~\cite{Liu2025simulationofopen,delgado2025quantum}, computational fluid dynamics~\cite{CHEN2024117428}, and finance~\cite{kumar2025simulating}. The query theorem applies to representations satisfying the stated input and dissipativity assumptions. 
	
Then applying the CBMD framework to simulate non-unitary dynamics described in Eq.~\eqref{eqn:ODE}, we set appropriate auxiliary functions $h_1(z)$ and $h_2(z)$ such that the integral remainder $\epsilon_{\text{IR}}$ vanishes. This effectively decomposes the non-Hermitian evolution into a linear combination of Hermitian evolutions. This conversion is formalized below, with proof provided in the Appendix~\ref{proof01}.

	\begin{lemma}[Matrix Series Identity]\label{Matrix_Series}
Let $A(t)=L(t)+iH(t)$ have bounded piecewise-continuous Hermitian parts on $[0,T]$, with $T\ge0$ and $L(t)\succeq0$ pointwise. Let $a>0$ satisfy $\int_0^T\norm{L(s)}\,ds\le2\pi a$, and let $\{p_r\}_{r=1}^m$, $m\ge1$, be distinct complex numbers avoiding $-i$ and $\mathbb Z/a$. Then the real-pole series is absolutely convergent and
\begin{align}\label{series_MSI}
&\mathcal T\e^{-\int_0^T A(s)\,ds}\nonumber\\
&-\frac1a\sum_{k\in\mathbb Z}
\frac{(\e^{-2\pi a}-1)\mathcal T\e^{-i\int_0^T(H(s)+(k/a)L(s))\,ds}}
{2\pi i(k/a+i)\displaystyle\prod_{r=1}^m\frac{k/a-p_r}{-i-p_r}}\nonumber\\
&-\sum_{r=1}^m
\frac{(\e^{-2\pi a}-1)\mathcal T\e^{-i\int_0^T(H(s)+p_rL(s))\,ds}}
{(\e^{-2\pi iap_r}-1)\displaystyle\prod_{\substack{r'=1\\r'\ne r}}^m\frac{p_r-p_{r'}}{-i-p_{r'}}}=0.
\end{align}
\end{lemma}
Appendix~\ref{proof01} proves the identity using the logarithmic-norm bound for time-ordered evolution. In general $\int_0^T\norm{L(s)}\,ds$ cannot be replaced by $\norm{\int_0^T L(s)\,ds}$.

\begin{lemma}[Fast-Converging LCU Series]\label{Query_Analyse}
Under the assumptions on $A(t)$ in Lemma~\ref{Matrix_Series}, for $0<\epsilon_1<1/2$ there exists an LCU $\sum_jc_jU_j$ approximating $\mathcal T\e^{-\int_0^T A(s)\,ds}$ within operator-norm error $\epsilon_1$, where
\begin{equation}
U_j=\mathcal T\e^{-i\int_0^T(H(s)+k'_jL(s))\,ds},\qquad k'_j\in\mathbb R.
\end{equation}
The number of distinct unitaries is
\begin{equation}
\mathcal O\left(\log\frac1{\epsilon_1}\left[\log\frac1{\epsilon_1}+\int_0^T\norm{L(s)}\,ds\right]\right),
\end{equation}
and
\begin{equation}
\max_j|k'_j|=\mathcal O\left(\log\frac1{\epsilon_1}\right),\qquad
\sum_j|c_j|=\mathcal O(1).
\end{equation}
\end{lemma}
The proof and the choices of auxiliary poles are given in Appendix~\ref{proof02}.

\begin{theorem}[Non-Hermitian simulation via CBMD]\label{thmhode}
Consider Eq.~\eqref{eqn:ODE} with $b(t)\equiv0$, $A(t)=L(t)+iH(t)$, bounded piecewise-continuous Hermitian $H(t),L(t)$ on $[0,T]$, and $L(t)\succeq0$ pointwise. The supplied matrix oracle block-encodes $A(t)/\alpha_A$ as in Eq.~\eqref{eq:input}, with inverse and controlled access. Write $u_T=u(T)\ne0$, and assume a constant-factor upper estimate of $\norm{u_0}/\norm{u_T}$ is supplied. For $0<\epsilon<1/2$, the normalized state at time $T$ can be prepared with vector-norm error $\epsilon$ and success probability $\Omega(1)$. The query complexity to the block-encoding of $A(t)$ is
\begin{equation}
\widetilde{\mathcal O}\left(\frac{\norm{u_0}}{\norm{u_T}}(1+\alpha_A T)
\log^2\frac{2\norm{u_0}}{\epsilon\norm{u_T}}\right)
\end{equation}
in the time-dependent case, assuming the coherent oracle and time-grid accuracy specified in Appendix~\ref{proof02}, with $\alpha_A\ge\sup_s\norm{A(s)}$. For time-independent $A$, it is
\begin{equation}\label{eq:firstquery}
\mathcal O\left(\frac{\norm{u_0}}{\norm{u_T}}(1+\alpha_A T)
\log\frac{2\norm{u_0}}{\epsilon\norm{u_T}}\right).
\end{equation}
In both cases, the query complexity to the state-preparation oracle of $|u_0\rangle$ is $\mathcal O(\norm{u_0}/\norm{u_T})$. The tilde suppresses logarithmic factors from Hamiltonian simulation. Time-index arithmetic contributes to gate cost only.
\end{theorem}
The proof is given in Appendix~\ref{proof02}.

\begin{table}[tb]
\caption{Sufficient sampling cutoffs and coefficient weights at a common \emph{operator} error $\epsilon_1$. For normalized-state preparation, substitute $\epsilon_1=\Theta(\epsilon\norm{u_T}/\norm{u_0})$ before including the $O(\norm{u_0}/\norm{u_T})$ amplification factor.}\label{tab:comparison}
\begin{tabular}{lll}\toprule
Construction & Real cutoff & Coefficient weight\\\midrule
Cauchy kernel~\cite{PhysRevLett.131.150603}&$O(1/\epsilon_1)$&$O(1)$\\
Fixed $0<\beta<1$~\cite{an2023quantumalgorithmlinearnonunitary}&$O_\beta(\log^{1/\beta}(1/\epsilon_1))$&$O_\beta(1)$\\
Gaussian kernel~\cite{low2025optimalquantumsimulationlinear}&$O(\log(1/\epsilon_1))$&$O(1)$\\
CBMD rational poles &$O(\log(1/\epsilon_1))$&$O(1)$\\\bottomrule
\end{tabular}
\end{table}
Appendix~\ref{lchs_conn} gives the corresponding discrete identities. At a common operator error, CBMD and the Gaussian LCHS kernel have the same logarithmic real cutoff and bounded coefficient weight.

\begin{figure}[tb]\centering
\includegraphics[width=\textwidth]{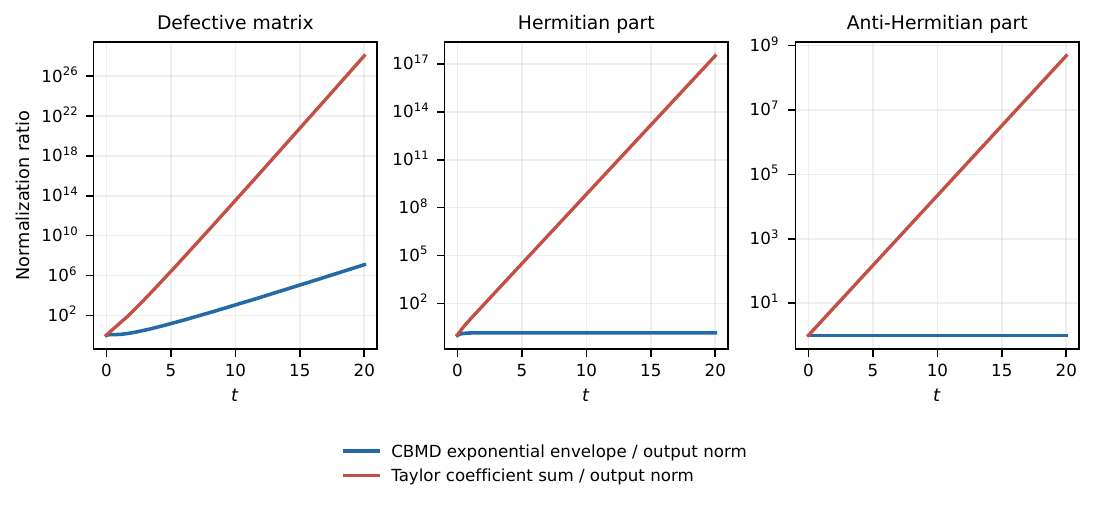}
\caption{First-order normalization comparison. Blue shows $\norm{u_0}/\norm{u(t)}$, the factor in CBMD up to its bounded coefficient weight. Red shows the direct Taylor--LCU normalization $\exp(\alpha_A t)\norm{u_0}/\norm{u(t)}$, with $\alpha_A$ taken as the norm of each panel's generator in the single-oracle model.}\label{fig1}
\end{figure}

To illustrate the normalization cost, consider the defective matrix
\begin{equation}
iH+L=\begin{pmatrix}1&2\\0&1\end{pmatrix}
\end{equation}
and the input $|u_0\rangle=(0,1)^T$. Under the full generator, its Hermitian part $L$, and its anti-Hermitian part $iH$, the output norms are $\e^{-t}\sqrt{1+4t^2}$, $\sqrt{(1+\e^{-4t})/2}$, and $1$, respectively. The corresponding generator norms are $1+\sqrt2$, $2$, and $1$. These values give the curves in Fig.~\ref{fig1}.

\section{Application II: Second-Order Wave Dynamics}\label{sec8}

We consider a non-Hermitian extension of the wave equation studied in Ref.~\cite{PhysRevA.99.012323}. Starting here, the matrix oracles for all three second-order problems (wave, Bessel $J_0$, and Airy) are separate block-encodings of $H/\norm H$ and $L/\norm L$. Both CBMD and the direct Taylor--LCU comparison use these same inputs.

Consider the dynamics of a physical system governed by a second-order wave equation with non-conservative terms. The evolution of the state $u(t)$ at an arbitrary time $t > 0$ can be described by:
\begin{equation}\label{eq:wave}
	\frac{d^2}{dt^2}u(t) = -A u(t), \quad u(0) = u_0, \quad u'(0) = 0,
\end{equation}
where $A = L + iH$ is a general non-Hermitian operator, with $L, H \in \mathbb{C}^{N \times N}$ being Hermitian matrices and {$L\succ 0, H\succ 0$}. The analytical solution at time $t$ is given by the matrix wave propagator:
\begin{equation} \label{eq:analytical_solution}
	u(t) = \cos\left(\sqrt{A}t\right) u_0 = \cos\left(\sqrt{L + iH}t\right) u_0.
\end{equation}
The power series $\cos(t\sqrt A)=\sum_{n\ge0}(-A)^nt^{2n}/(2n)!$ converges for every matrix. Termwise differentiation verifies Eq.~\eqref{eq:wave} and both initial conditions. Our objective is to prepare a normalized quantum state $|u(t)\rangle$ with a success probability of $\Omega(1)$ and within a precision of $\epsilon$.

To decompose this non-Hermitian function into an LCU composed of purely Hermitian arguments, we construct the following complex-valued auxiliary function parameterized by $t$:
\begin{equation}
	F(z) = \frac{h(1) \cos\left(\sqrt{izH + L} \ t\right)}{h(z) \cdot z(z-1)},
\end{equation}
where $h(z) = \cos\left(\sqrt{2\pi a i z} \ t\right)$.

The contour estimate in Appendix~\ref{proofthm5} holds for $2\pi a\ge\norm H$. For the query bounds we take $2\pi a=\norm H$. The function $h(z)$ is an entire function, and its reciprocal yields strictly simple poles. The proof uses circles on which $|\cos(t\sqrt{2\pi a i z})|$ has a uniform angular lower bound, also for noncommuting $L,H$. Applying Cauchy's residue theorem along this sequence, the sum of all residues vanishes: $\sum \text{Res}(F, z_p) = 0$. By combining all the first-order poles, we arrive at the following exact LCU identity, where the truncation error terms are explicitly distinguished:

\begin{equation} \label{eq:lcu_exact}
	\begin{split}
		\cos\left(\sqrt{A}t\right) = & \underbrace{h(1) \cos\left(\sqrt{L}t\right) -  h(1) \sum_{k=0}^K d_k \cos\left(\sqrt{\frac{(k+1/2)^2\pi}{2a t^2}H + L} \ t\right)}_{\text{Quantum Circuit Implementation by QSVT+LCU}} \\
		& -  \underbrace{h(1) \sum_{k=K+1}^\infty d_k \cos\left(\sqrt{\frac{(k+1/2)^2\pi}{2a t^2}H + L} \ t\right)}_{\text{Truncation Error}},
	\end{split}
\end{equation}
with coefficients
\begin{equation}\label{eq:regulator}
	d_k = \frac{(-1)^k}{\frac{(k+1/2)\pi}{2} \left( 1 +  i \frac{(k+1/2)^2\pi}{2a t^2} \right)}.
\end{equation}

The two block-encodings are exact, with controlled and inverse access. Both norms are known. The input-state unitary prepares $u_0/\norm{u_0}$ and its inverse is available. For every $s\ge0$, a two-term LCU yields
\begin{equation}\label{eq:separate-oracles}
\frac{L+sH}{\norm L+s\norm H}
=\frac{\norm L}{\norm L+s\norm H}\frac{L}{\norm L}
+\frac{s\norm H}{\norm L+s\norm H}\frac{H}{\norm H}.
\end{equation}
Each use costs $O(1)$ calls to the two supplied oracles. Its normalized spectrum is contained in $[1/\kappa,1]$, with the original condition parameter
\begin{equation}\label{eq:kappabe}
\kappa=\max(\kappa_H,\kappa_L),\qquad
\kappa_H=\frac{\norm H}{\lambda_{\min}(H)},\quad
\kappa_L=\frac{\norm L}{\lambda_{\min}(L)}.
\end{equation}
We assume known positive lower bounds on $\lambda_{\min}(H)$ and $\lambda_{\min}(L)$, each within a fixed factor of its eigenvalue. Substituting those bounds in Eq.~\eqref{eq:kappabe} gives a computable upper bound on $\kappa$ within a fixed factor. With looser spectral promises the query bounds use this computable upper bound. All query statements exclude coefficient-loading, reversible-arithmetic and phase-synthesis gate costs.

Based on the preceding analysis, we establish the following theorem (proof provided in Appendix~\ref{proofthm5}):

\begin{theorem}[Complexity of Non-Hermitian Wave Equation]\label{wavethm}
Consider Eq.~\eqref{eq:wave} with $A=L+iH$, $L,H\succ0$, and separate input block-encodings of $H/\norm H$ and $L/\norm L$ as in Eq.~\eqref{eq:separate-oracles}. Suppose $t>0$, $0<\epsilon<1/2$, and a positive lower bound on $\norm{u(t)}/\norm{u_0}$ is given within a fixed factor of this ratio. To prepare the normalized state with vector-norm error $\epsilon$ and success probability $\Omega(1)$, the total number of calls to the two matrix oracles and the number of input-state calls are, respectively,
\begin{equation}\label{eq:special-cost}
\widetilde{\mathcal O}\left(u_r\kappa\left[1+t\sqrt{\norm L+\frac{u_r\norm H}{\epsilon}}\right]\right)
\quad\text{and}\quad
\mathcal O\left(u_r\log(2+\norm H\,t^2)\right),
\end{equation}
where the amplification factor is
\begin{equation}\label{eq:wave-ur}
u_r:=\frac{\norm{u_0}\e^{\sqrt{\norm H/2}\,t}}{\norm{\cos(\sqrt A t)u_0}},
\qquad \kappa=\max(\kappa_H,\kappa_L).
\end{equation}
The tilde hides only logarithmic factors in the displayed parameters. The case $t=0$ is the identity.
\end{theorem}

For the stated choice $2\pi a=\norm H$, the CBMD block-encoding normalization is bounded by
\begin{equation}\label{eq:envelope}
O\left(e^{t\sqrt{\norm H/2}}\log(2+\norm H\,t^2)\right).
\end{equation}
This is a valid upper bound and need not equal $\norm{\cos(\sqrt A t)}$. The factor $u_r$ can be favorable when the output norm grows with the same envelope, but is not uniformly constant.

For comparison, the same separate oracles give a block-encoding of $A/(\norm L+\norm H)$ with $O(1)$ queries. A direct Taylor--LCU construction then has coefficient sum $\cosh(t\sqrt{\norm L+\norm H})$. CBMD reduces the exponential normalization envelope to one governed by $\norm H$. Its full query bound still contains $t\sqrt{\norm L}$ and the truncation term proportional to $u_r^{3/2}/\sqrt\epsilon$. Thus the improvement can survive when $L$ is large compared with $H$, conditioning is moderate and the output is not too small, but it is not uniform in precision or output norm. Appendix~\ref{proofthm5} compares the complete bounds in this same input model.

\textbf{Remark.} The $\norm L$ term cannot simply be omitted: even for a small anti-Hermitian component, the algorithm must implement the oscillatory Hermitian transformation $\cos(t\sqrt L)$. Separate normalized input oracles restore $\kappa=\max(\kappa_H,\kappa_L)$; they do not make this transformation free.

\begin{figure}[H] 
    \centering 
    \includegraphics[width=1.0\textwidth]{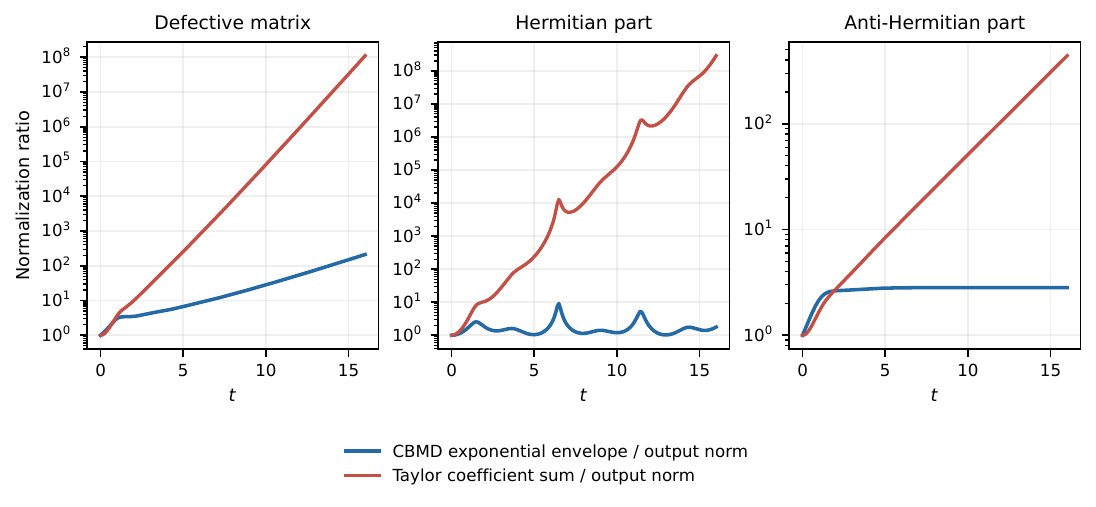} 
    \caption{Wave normalization comparison for the matrix in Eq.~\eqref{eq:benchmarkA} and its Hermitian and anti-Hermitian parts. Blue is $e^{t\sqrt{\norm H/2}}\norm{u_0}/\norm{u(t)}$; red is $\cosh(t\sqrt{\norm L+\norm H})\norm{u_0}/\norm{u(t)}$. Blue omits logarithmic coefficient and fixed QSVT factors; these are not complete query counts.} 
    \label{fig2}
\end{figure}

In the case of the wave function, we compare normalization envelopes for the defective matrix
\begin{equation}\label{eq:benchmarkA}
iH+L=\begin{pmatrix}1+i&1\\0&1+i\end{pmatrix}.
\end{equation}
Both Hermitian parts have eigenvalues $1/2,3/2$. We use $|u_0\rangle=(0,1)^T$ and the same input for Bessel and Airy. Figure~\ref{fig2} compares the CBMD envelope with the direct Taylor coefficient sum under the separate-oracle input model. In Figs.~\ref{fig2}--\ref{fig4} the purely Hermitian and anti-Hermitian panels plot analytic limiting cases. A vanishing component contributes zero weight and is omitted. The query theorems assume $H,L\succ0$.

\section{Application III: Non-Hermitian Special Functions}\label{sec:spfunc}
\subsection{Non-Hermitian Bessel \texorpdfstring{$J_0$}{J0} Dynamics}

To demonstrate the versatility of the CBMD framework, we introduce a non-Hermitian dynamical differential equation governed by Bessel functions. Consider the matrix differential equation
\begin{equation}\label{eq:bessel}
	\frac{d^2u(t)}{dt^2} + \frac{1}{t} \frac{du(t)}{dt} + A u(t) = 0,
\end{equation}
for $t>0$, with the regular initial conditions $u(0)=u_0$ and $u'(0)=0$ at the singular endpoint. Here, $A = L +  i H$ represents a general non-Hermitian operator, where $L, H \in \mathbb{C}^{N \times N}$ are positive-definite Hermitian matrices ({$L\succ 0, H\succ 0$}). 

The term $t^{-1}u'(t)$ specifies a time-dependent damping coefficient in this mathematical model. Regularity at $t=0$ selects the Bessel solution; no microscopic thermal-bath model is assumed.

The regular solution is $u(t)=J_0(t\sqrt{A})u_0$, where
\begin{equation}\label{eq:bessel-propagator}
	J_0(t\sqrt A)=\sum_{m=0}^{\infty}\frac{(-A)^m t^{2m}}{4^m(m!)^2}.
\end{equation}
Recognizing that $J_0(\sqrt{z})$ is an entire function, our algorithmic goal is to prepare the normalized final quantum state $|u(t)\rangle \propto J_0(\sqrt{A}t)|u_0\rangle$ with a success probability of $\Omega(1)$.

To enable the direct evaluation of this non-Hermitian operator function, we must systematically suppress its exponential divergence in the complex plane. We use the same regulator $h(z)=\cos(t\sqrt{2\pi a i z})$, taking $2\pi a=\norm H$ for the query bound. By constructing the objective function
\begin{equation}
	F(z) = \frac{h(1) J_0(\sqrt{ i zH+L}t)}{h(z) z(z-1)},
\end{equation}
and evaluating the pole-avoiding contour sequence of Appendix~\ref{proofthm5}, we obtain an exact series expansion for the target dynamics:
\begin{equation}
	J_0(\sqrt{A}t) = h(1) J_0(\sqrt{L}t) - h(1) \sum_{k=0}^{\infty} d_k J_0\left( \sqrt{\frac{(k + 1/2)^2 \pi}{2a t^2} H + L} t \right),
\end{equation}
where the expansion coefficients are given by
\begin{equation}
	d_k = \frac{(-1)^k}{\frac{(k + 1/2)\pi}{2} \left( 1 +  i  \frac{(k + 1/2)^2 \pi}{2a t^2} \right)}.
\end{equation}
This expansion decomposes the target into functions of positive Hermitian matrices. The identity
\begin{equation}\label{eq:bessel-cos}
J_0(t\sqrt X)=\frac1\pi\int_0^\pi\cos(t\cos\theta\sqrt X)\,d\theta
\end{equation}
implies $|J_0(t\sqrt x)|\le1$ for $x\ge0$ and allows the same contour and QSVT bounds as for the wave equation~\cite{NISTDLMF}. The following theorem uses the separate normalized $H,L$ oracles of Application II; its proof is in Appendix~\ref{besselproof}.

\begin{theorem}[Complexity of Non-Hermitian Bessel Equation]\label{besselthm}
Consider Eq.~\eqref{eq:bessel} with $A=L+iH$, $L,H\succ0$, and separate input block-encodings of $H/\norm H$ and $L/\norm L$ as in Eq.~\eqref{eq:separate-oracles}. Suppose $t>0$, $0<\epsilon<1/2$, and a positive lower bound on $\norm{u(t)}/\norm{u_0}$ is given within a fixed factor of this ratio. To prepare the normalized state with vector-norm error $\epsilon$ and success probability $\Omega(1)$, the total number of calls to the two matrix oracles and the number of input-state calls are, respectively,
\begin{equation}
\widetilde{\mathcal O}\left(u_r\kappa\left[1+t\sqrt{\norm L+\frac{u_r\norm H}{\epsilon}}\right]\right)
\quad\text{and}\quad
\mathcal O\left(u_r\log(2+\norm H\,t^2)\right),
\end{equation}
where the amplification factor is
\begin{equation}
u_r:=\frac{\norm{u_0}\e^{\sqrt{\norm H/2}\,t}}{\norm{J_0(\sqrt A t)u_0}},
\qquad \kappa=\max(\kappa_H,\kappa_L).
\end{equation}
The tilde hides only logarithmic factors in the displayed parameters. The case $t=0$ is the identity.
\end{theorem}

\begin{figure}[H] 
    \centering 
    \includegraphics[width=1.0\textwidth]{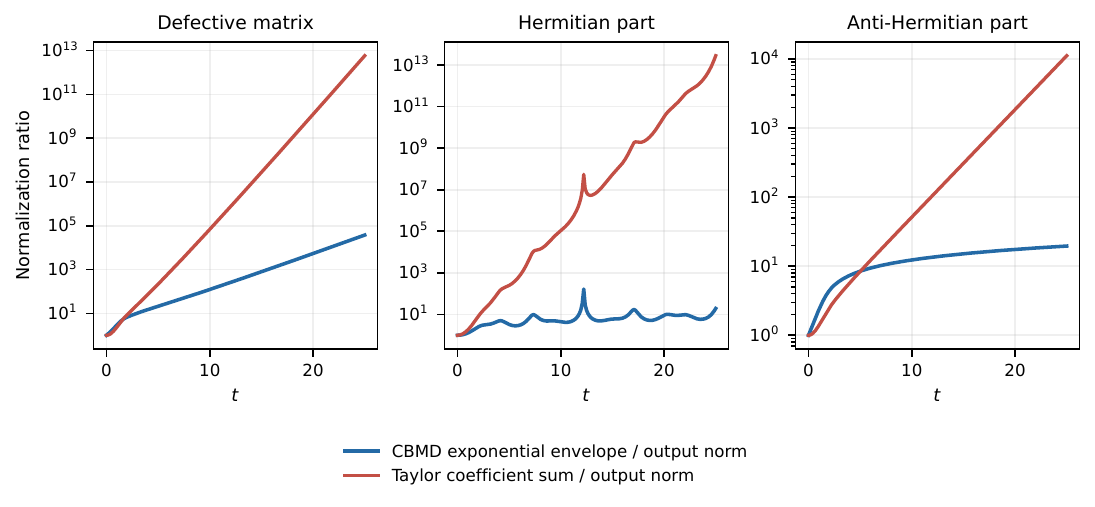} 
    \caption{Bessel normalization comparison for the same inputs as Fig.~\ref{fig2}. Blue is $e^{t\sqrt{\norm H/2}}\norm{u_0}/\norm{u(t)}$ and red is $I_0(t\sqrt{\norm L+\norm H})\norm{u_0}/\norm{u(t)}$, with the same qualification concerning omitted factors.} 
    \label{fig3}
\end{figure}

For Bessel dynamics we use the same matrix and input as in Eq.~\eqref{eq:benchmarkA}, and compare normalization envelopes in Fig.~\ref{fig3}. The direct Taylor coefficient sum is $I_0(t\sqrt{\norm L+\norm H})$. The numerator $e^{t\sqrt{\norm H/2}}$ of $u_r$ controls the CBMD coefficient envelope, while the output norm and the polynomial degree also affect the full complexity. For the scalar boundary case $A=\ii h$ with fixed $h>0$, the envelope ratio grows as $\Theta(t^{1/2})$ as $t\to\infty$.

\subsection{Non-Hermitian Airy Dynamics via \texorpdfstring{$\mathbb{Z}_3$}{Z3} Symmetry Projection}

As another {mathematical test case for} higher-order non-Hermitian evolution, we consider {an idealized model of} a parametric oscillator with a linearly varying generalized stiffness governed by the Airy equation:
\begin{equation}\label{eq:airy_new}
	\frac{d^2u(t)}{dt^2} + t A u(t) = 0,
\end{equation}
subject to the initial conditions $u(0)=u_0$ and $u'(0)=0$. We use this as a mathematical model of a linearly varying stiffness. A single Airy branch does not in general satisfy both prescribed initial conditions. Although the scalar Airy function has order $3/2$, the projected solution below is an entire function of the matrix argument; this is the relevant property for the contour construction.

To impose the initial conditions, we introduce the $\mathbb{Z}_3$ symmetry projection
\begin{equation}\label{eq:airy-projection}
	f(z) = \frac{1}{3} \left( \mathrm{Ai}(-z) + \mathrm{Ai}(-\omega z) + \mathrm{Ai}(-\omega^2 z) \right),
\end{equation}
where $\omega=e^{2\pi i/3}$ and $f(0)=\Ai(0)$. The root-of-unity projection retains exactly the powers $z^{3k}$. Therefore $f(A^{1/3}t)$ is defined by an entire series in $A$, without computing a matrix cube root. The scalar function $f(z)$ still has order $3/2$ in $z$; for fixed $t>0$, its composition has order $1/2$ in the matrix argument. The solution with the prescribed initial value is
\begin{equation}\label{eq:airy-entire}
u(t)=\frac{f(A^{1/3}t)}{f(0)}u_0.
\end{equation}
Appendix~\ref{proof_airy} verifies both initial conditions and the differential equation.

To evaluate this target dynamics using the methodology established in the previous section, we define a trigonometric asymptotic regulator $h(z) = \cos\left(\frac{2}{3} t^{3/2}\sqrt{ i \gamma z}\right)$. The $2/3$ scaling follows from the Airy--Bessel integral representation in Appendix~\ref{proof_airy}. The contour identity permits $\gamma\ge\norm H$; to obtain the complexity below we choose $\gamma=\norm H$ and construct the analogous objective function
\begin{equation}
	F(z) = \frac{h(1) f(( i zH+L)^{1/3}t)}{h(z) z(z-1)},
\end{equation}
whose contour integral vanishes along the sequence in Appendix~\ref{proofthm5}. The simple zeros of $h$ are the poles of $1/h$. They occur at $z_k=-i\,9(k+1/2)^2\pi^2/(4\gamma t^3)$ for $k\ge0$. At these points the argument $iz_kH+L$ is positive Hermitian. Summing the residues gives the absolutely convergent identity
\begin{equation}
	f(( i H+L)^{1/3}t) = h(1) f(L^{1/3}t) - h(1) \sum_{k=0}^{\infty} d_k f\left( \left( L + \frac{9(k+1/2)^2\pi^2}{4\gamma t^3} H \right)^{1/3} t \right),
\end{equation}
where $h(0)=1$ and the residue coefficients are
\begin{equation}
	d_k = \frac{1}{h'(z_k) z_k (z_k - 1)} = \frac{(-1)^k}{\frac{(k+1/2)\pi}{2} \left( 1 +  i \frac{9(k+1/2)^2\pi^2}{4\gamma t^3} \right)}.
\end{equation}

After division by $f(0)$, the positive-real-axis function is a probability average of cosines and is bounded by one. The same separate $H,L$ oracles and QSVT construction therefore apply. The following theorem is proved in Appendix~\ref{proof_airy}.

\begin{theorem}[Complexity of Non-Hermitian Airy Dynamics via $\mathbb Z_3$ Projection]\label{airythm}
Consider Eq.~\eqref{eq:airy_new} with $A=L+iH$, $L,H\succ0$, and separate input block-encodings of $H/\norm H$ and $L/\norm L$ as in Eq.~\eqref{eq:separate-oracles}. Suppose $t>0$, $0<\epsilon<1/2$, and a positive lower bound on $\norm{u(t)}/\norm{u_0}$ is given within a fixed factor of this ratio. The normalized state can be prepared with vector-norm error $\epsilon$ and success probability $\Omega(1)$ using, respectively, the following total queries to the two matrix oracles and queries to the input-state oracle:
\begin{equation}\label{eq:airy-complexity}
\widetilde{\mathcal O}\left(u_r\kappa\left[1+t^{3/2}\sqrt{\norm L+\frac{u_r\norm H}{\epsilon}}\right]\right)
\quad\text{and}\quad
\mathcal O\left(u_r\log(2+\norm H\,t^3)\right),
\end{equation}
where
\begin{equation}\label{eq:airy-ur}
u_r:=\frac{\norm{u_0}\exp\left(\frac23\sqrt{\frac{\norm H}{2}t^3}\right)}{\norm{f(A^{1/3}t)u_0}},
\qquad \kappa=\max(\kappa_H,\kappa_L).
\end{equation}
The tilde hides logarithmic factors only. The truncation contribution retains its $1/\sqrt\epsilon$ dependence. The case $t=0$ is the identity.
\end{theorem}

The fixed factor $f(0)=\Ai(0)$ is absorbed into the constants in the query bounds; the propagator remains $f(A^{1/3}t)/f(0)$. Thus $u_r$ retains its original definition. Appendix~\ref{proof_airy} tracks this normalization explicitly.

\begin{figure}[H] 
    \centering 
    \includegraphics[width=1.0\textwidth]{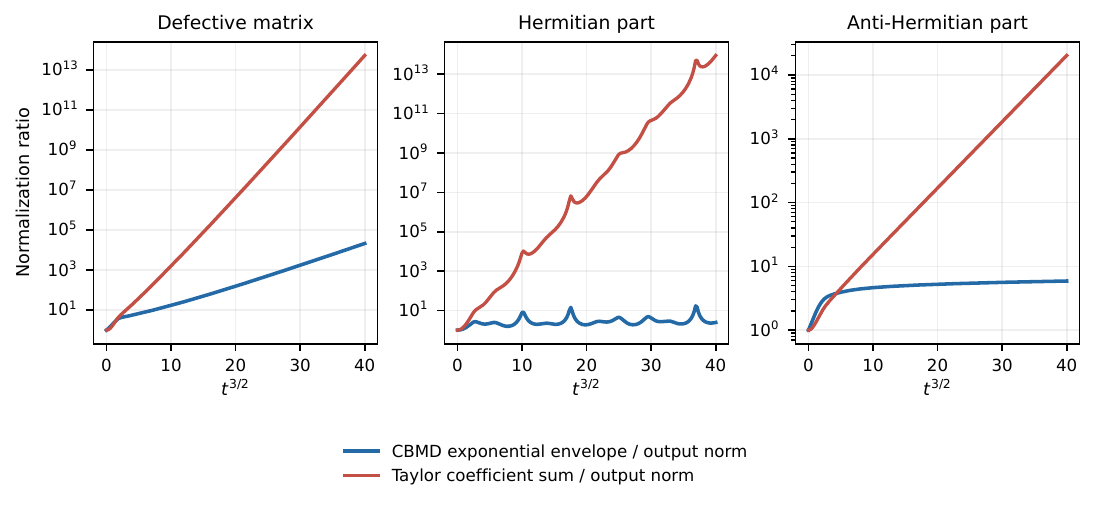} 
    \caption{Airy normalization comparison versus $t^{3/2}$. Blue is $e^{(2/3)t^{3/2}\sqrt{\norm H/2}}\norm{u_0}/\norm{u(t)}$; red is ${}_0F_1(;2/3;(\norm L+\norm H)t^3/9)\norm{u_0}/\norm{u(t)}$. The blue ratio equals $|f(0)|u_r$. The initial value includes division by $f(0)$. These curves omit truncation and polynomial-degree costs.} 
    \label{fig4}
\end{figure}

For Airy dynamics we use the same matrix and input as in Eq.~\eqref{eq:benchmarkA}; Fig.~\ref{fig4} compares the normalized propagator with the direct Taylor coefficient sum ${}_0F_1(;2/3;(\norm L+\norm H)t^3/9)$. The state norm and the polynomial degree must be included in a full cost comparison. For the scalar boundary case $A=\ii h$ with fixed $h>0$, the envelope ratio grows as $\Theta(t^{1/4})$ as $t\to\infty$.

\section{Discussion}

The auxiliary function $h(z)$ compensates for growth on the complex contours. For wave and Bessel dynamics its trigonometric factors give an absolute-sum bound on the residue coefficients. This can reduce normalization relative to a direct Taylor--LCU construction, while the total cost retains the dependence on spectral gaps, precision and output norm stated in the theorems.

For the Airy initial-value problem, the $\mathbb Z_3$ projection retains the powers divisible by three. Dividing by $f(0)$ gives an entire propagator with the prescribed initial data. An integral of cosines against a positive measure then gives the bounds required by the same residue argument.

For wave, Bessel $J_0$, and Airy dynamics, separate input block-encodings of $H/\norm H$ and $L/\norm L$ give the spectral condition parameter $\max(\kappa_H,\kappa_L)$. The query bounds also include the cost of transforming $L$. Consider fixed $t>0$ and bounded $\norm H$ as $\norm L$ increases. If $\max(\kappa_H,\kappa_L)$, $\norm{u_0}/\norm{u(t)}$ and $1/\epsilon$ are at most polynomial in $\norm L$, the CBMD query bounds are polynomial in $\norm L$. In this regime the specified direct Taylor coefficient sums are exponential in $\sqrt{\norm L+\norm H}$. This comparison applies to those monomial constructions. Conversely, at fixed matrices and time, the direct Taylor truncation admits logarithmic precision dependence while the present CBMD bound has algebraic precision dependence.

One open question is whether higher-order poles give faster residue convergence at comparable coefficient weight. Another is whether scalar rational approximation, such as the AAA algorithm~\cite{doi:10.1137/16M1106122}, can guide the choice of auxiliary poles with provable truncation and weight bounds.

\section*{Acknowledgements}
	This work has been supported by the National Key Research and Development Program of China (Grant No. 2023YFB4502500). 
\section*{Declarations}
\begin{itemize}
	\item The authors declare that they have no conflict of interest.
	\item Reproducible figure data and numerical verification scripts accompany this revision. They check finite examples and do not replace the analytic proofs.
\end{itemize}

\appendix
\section{Proof of Matrix Series Identity Lemma~\ref{Matrix_Series}}\label{proof01}
For the proofs only, abbreviate
\begin{equation}\label{eq:rho}
V=\mathcal T\e^{-\int_0^T A(s)ds},\qquad
U(z)=\mathcal T\e^{-i\int_0^T(H(s)+zL(s))ds},\qquad
\rho_L=\int_0^T\norm{L(s)}ds.
\end{equation}
Then $V=U(-i)$ and $\norm V\le1$. We also write $|\psi_0\rangle=u_0/\norm{u_0}$ in the proofs. Define
\begin{equation}
P(z)=\prod_{r=1}^m\frac{z-p_r}{-i-p_r},\qquad
w_k=\frac{\e^{-2\pi a}-1}{2\pi ia(k/a+i)P(k/a)},
\end{equation}
and
\begin{equation}
R_r=-\frac{(\e^{-2\pi a}-1)U(p_r)}
{(\e^{-2\pi iap_r}-1)\displaystyle\prod_{j\ne r}(p_r-p_j)/(-i-p_j)}.
\end{equation}
These abbreviations are not additional inputs to Lemma~\ref{Matrix_Series}.

For $z=x+\ii y$, the Hermitian part of the generator $-\ii H(t)-\ii zL(t)$ is $yL(t)$. If $v(t)$ solves the corresponding evolution, differentiating $\norm{v(t)}^2$ gives
\begin{equation}\label{norm_len1}
\norm{U(x+\ii y)}\le
\begin{cases}\exp(y\rho_L),&y\ge0,\\1,&y<0.\end{cases}
\end{equation}
Pointwise $L(t)\succeq0$ is used here. The Dyson series converges uniformly on compact $z$-sets, so $U(z)$ is entire. A bound involving $\norm{\int L}$ is not justified by this argument for noncommuting time-dependent evolution.

For example, use $L_1=\operatorname{diag}(1,0)$ for one unit interval, then a unitary swap generated in an interval with $L=0$, then $L_2=\operatorname{diag}(0,1)$ for another unit interval. For positive real $q$, the resulting growing evolution is, up to phase,
$\e^{qL_2}\sigma_x\e^{qL_1}$, with norm $\e^{2q}$. Here $\norm{\int L}=\norm{L_1+L_2}=1$, whereas $\rho_L=2$. To see why pointwise dissipativity matters as well, keep the swap and instead take $L_1=\operatorname{diag}(-1,1)$ and $L_2=-L_1$. Then $\int L=0$, but $\norm{\e^{-L_2}\sigma_x\e^{-L_1}}=\e^2$.

Let
\begin{equation}\label{int_main}
F(z)=\frac{(\e^{-2\pi a}-1)U(z)}{(z+\ii)P(z)(\e^{-2\pi\ii az}-1)}.
\end{equation}

		\begin{figure}[htbp]
			\centering 
			\begin{tikzpicture}[
				font=\small,
				>={Stealth[length=1mm, width=1mm]},declare function={
					R=3.5;
					r=0.05;
					offset=0.05;
				}
				]
				
				\draw[gray, ->, thin] (-R-0.5,0) -- (R+0.5,0) node[below left] {Re};
				\draw[gray,->, thin] (0,-R-0.5) -- (0,R+0.5) node[below left] {Im};
				\filldraw (0,0) circle (0.8pt) node[below right] {$O$};
				\draw[blue, -, semithick] (-R, -R) -- (-R, 0.8);
				\draw[blue,<-, semithick] (-R, 0.7) -- (-R, R+0.012);
				
				\draw[blue,-, semithick] (-R, R) -- (0.8, R);
				\draw[blue,<-, semithick] (0.7, R) -- (0.012+R, R+0.012);
				
				\draw[blue,-, semithick] (R, R) -- (R, -0.8);
				\draw[blue,<-, semithick] (R, -0.7) -- (R, -R-0.012);
				
				\draw[blue,-, semithick] (R, -R) -- (-0.8, -R);
				\draw[blue,<-, semithick] (-0.7, -R) -- (-R, -R-0.012);

				\node[blue] at (-R*0.85, R*0.85) {$\Gamma$};
				\node[black,fill=white,inner sep=1.5pt] at (0, R*0.35) {$\frac{1}{2\pi i }\oint_{\color{blue}\Gamma} \frac{({\mathrm{e}^{-2\pi a}-1})\mathcal{T}\mathrm{e}^{- i \int_{0}^{T}(H(s)+zL(s))\mathrm{d}s}}{(z+ i )\left[
						\prod_{r=1}^{m}\frac{z-p_{r}}{- i -p_{r}}
						\right](\mathrm{e}^{-2\pi  i az}-1)}\mathrm{d}z$};
				\node[blue] at (R+0.3, R+0.3) {$+R+ i R$};
				\node[blue] at (-R-0.3, R+0.3) {$-R+ i R$};
				\node[blue] at (R+0.3, -R-0.3) {$+R- i R$};
				\node[blue] at (-R-0.3, -R-0.3) {$-R- i R$};		
			\end{tikzpicture}
			\caption{Schematic diagram of the matrix contour integral in Eq.~\eqref{int_main}.}
			\label{fig:tikz_rectangle_01}
		\end{figure}
Take positively oriented squares with half-width
$R_n=(n+1/2)/a>\max\{1,|p_1|,\ldots,|p_m|\}$. On a vertical side $x=\pm R_n$,
\begin{equation}
|\e^{-2\pi\ii az}-1|=1+\e^{2\pi ay},\qquad
\frac{\norm{U(z)}}{1+\e^{2\pi ay}}\le1,
\end{equation}
by $2\pi a\ge\rho_L$. On the top side the ratio $\norm{U(z)}/|\e^{-2\pi\ii az}-1|$ is at most $\e^{\rho_L R_n}/(\e^{2\pi a R_n}-1)$, and on the bottom it is at most $(1-\e^{-2\pi a R_n})^{-1}$. Both are bounded as $n\to\infty$. Uniformly on the boundary, $|(z+\ii)P(z)|^{-1}=O(R_n^{-m-1})$. The contour length is $8R_n$, so
\begin{equation}
\left\|\oint_{\Gamma_n}F(z)\,dz\right\|=O(R_n^{-m})\longrightarrow0.
\end{equation}

The residue at $-\ii$ is $V$. At a real pole $k/a$, the derivative of $\e^{-2\pi\ii az}-1$ is $-2\pi\ii a$, so the residue is $-w_kU(k/a)$. At $p_r$, differentiation of the single vanishing factor of $P$ gives exactly $R_r$ as defined above; the minus sign uses $(-\ii-p_r)/(p_r+\ii)=-1$. Consequently,
\begin{equation}
0=V-\sum_{k\in\mathbb Z}w_kU(k/a)+\sum_rR_r.
\end{equation}
Since $U(k/a)$ is unitary and $w_k=O(|k|^{-m-1})$, the real-pole sum converges absolutely. This proves Lemma~\ref{Matrix_Series}.

\section{Proof of Fast-Converging LCU Series Lemma~\ref{Query_Analyse}}\label{proof02}
Write $\delta=\epsilon_1$ in this proof. Choose the $2m+2$ auxiliary poles
$\{2i\}\cup\{r+i:-m\le r\le m\}$, with
\begin{equation}
m=\Theta(\log(C/\delta)),\qquad
a=\max\left\{1,\frac{\rho_L+\log(C(m+1)/\delta)}{2\pi}\right\},\qquad
K=\lceil2am\rceil.
\end{equation}
Here $m$ is a positive integer, and $C$ and the implicit lower constant in $m$ are sufficiently large universal constants. Put $x=k/a$. Euler's product gives, for fixed $c>0$,
\begin{equation}\label{eq:euler-product}
\prod_{r=1}^m(r^2+c^2)\le(m!)^2\frac{\sinh(\pi c)}{\pi c}.
\end{equation}
Throughout this appendix, constants are universal and can be increased independently of $m$, $a$, $\rho_L$, and $\delta$.

\subsection{Real-pole tail}
For $x\ge2m$,
\begin{equation}
|w_k|\le\frac{C(m!)^2}{a(x-m)^{2m+3}}.
\end{equation}
Indeed the coefficient numerator contains $6\prod_{r=1}^m(r^2+4)$, while each of the $2m+1$ factors $|x-r-\ii|$ and the two factors $|x+\ii|,|x-2\ii|$ is bounded below by $x-m$. Both tails are therefore bounded, using monotonicity and Stirling's inequality, by
\begin{equation}\label{eq:first-tail}
\sum_{|k|>K}|w_k|
\le\frac{C(m!)^2}{(K/a-m)^{2m+2}}
\le C\e^{-2m},\qquad K\ge2am,\quad m\ge1.
\end{equation}
Here $|\e^{-2\pi a}-1|\le1$, and $|w_{-k}|=|w_k|$ by symmetry of the auxiliary poles about the imaginary axis.

\subsection{Auxiliary residues}
At $p=2\ii$, the reciprocal of the absolute pole product in $R_{2\ii}$ is $2\prod_{r=1}^m(r^2+4)/(r^2+1)=O(1)$ by Eq.~\eqref{eq:euler-product}. Thus Eq.~\eqref{norm_len1} yields
\begin{equation}
\norm{R_{2\ii}}\le C\e^{2\rho_L-4\pi a}.
\end{equation}
For the poles $p=r+\ii$, $-m\le r\le m$, the product of integer separations contributes $(m-r)!(m+r)!$. Dropping the remaining denominators only increases the bound, and hence
\begin{align}
\sum_{r=-m}^m\norm{R_{r+\ii}}
&\le C\e^{\rho_L-2\pi a}(m!)^2
\sum_{r=-m}^m\frac1{(m-r)!(m+r)!}\\
&=C\e^{\rho_L-2\pi a}\frac{(m!)^2 4^m}{(2m)!}
\le C\sqrt{m+1}\e^{\rho_L-2\pi a}.
\end{align}
Here $a\ge1$ absorbs the constants from $(\e^{2\pi a}-1)^{-1}$. Choosing $m$ and $a$ as above, with a sufficiently large universal constant, makes the sum of these two bounds and Eq.~\eqref{eq:first-tail} at most $\delta$.

\subsection{Total coefficient weight}
For all real $x$,
\begin{equation}\label{eq:product-lower}
\prod_{r=-m}^m|x-r-\ii|\ge(m!)^2.
\end{equation}
The product is even. For $x\ge0$ its value at $x+1$ divided by its value at $x$ is $|x+m+1-\ii|/|x-m-\ii|\ge1$. It suffices to consider $0\le x<1$, where $|x-\ii|\ge1$ and
\begin{equation}
|x+r-\ii|^2 |x-r-\ii|^2
=r^4+2r^2(1-x^2)+(1+x^2)^2\ge r^4 .
\end{equation}
Multiplying the paired bounds proves Eq.~\eqref{eq:product-lower}. Combining it with Eq.~\eqref{eq:euler-product} gives
\begin{equation}
|w_k|\le\frac{C}{a\sqrt{1+x^2}\sqrt{4+x^2}}
\le\frac{C}{a(1+x^2)}.
\end{equation}
Finally,
\begin{equation}
\frac1a\sum_{k\in\mathbb Z}\frac1{1+(k/a)^2}
=\pi\coth(\pi a)=O(1),\qquad a\ge1.
\end{equation}
This proves the coefficient-weight claim. Relabel the retained $w_k,U(k/a),k/a$ as $c_j,U_j,k'_j$ for $|k|\le K$. Since $K=\lceil2am\rceil$, the term count and cutoff in Lemma~\ref{Query_Analyse} follow. No eigenvector conditioning enters this argument.

\subsection{Normalized-state preparation and Theorem~\ref{thmhode}}
For $T=0$ return the input state. For $T>0$ let $r$ be the supplied upper estimate, so $\norm{u_0}/\norm{u_T}\le r=O(\norm{u_0}/\norm{u_T})$, and put $\ell=\log(2r/\eps)$. Pointwise dissipativity gives $r\ge1$. Use $\epsilon_1=\eps/(8r)$ in Lemma~\ref{Query_Analyse}. To choose $a$, replace $\rho_L$ in its formula by the known bound $\alpha_A T$, since $L=(A+A^\dagger)/2$. This preserves the error and weight estimates and gives $K/a=O(\ell)$. The selected Hermitian generator is obtained from the original $A(t)/\alpha_A$ block-encoding and its adjoint by
\begin{equation}\label{eq:first-generator}
H(t)+xL(t)=\frac{x-i}{2}A(t)+\frac{x+i}{2}A(t)^\dagger,
\qquad \alpha_A\sqrt{1+x^2},\quad x\in\mathbb R,
\end{equation}
where the second expression is its block-encoding normalization. A controlled scalar rotation pads each branch to the common normalization $\Lambda=\alpha_A\sqrt{1+(K/a)^2}=O(\alpha_A\ell)$. This uses $O(1)$ controlled calls to the input oracle or its inverse, also for a coherent sampling index.

In the time-dependent case, the supplied grid must satisfy
\begin{equation}
\sum_{|k|\le K}|w_k|\norm{U(k/a)-U_{\rm grid}(k/a)}\le\frac{\eps}{16r}.
\end{equation}
Here $U_{\rm grid}$ evolves under the piecewise-constant interpolant on the supplied grid. To simulate it, divide $[0,T]$ into $O(1+\Lambda T)$ segments of length $d$ with $\Lambda d\le1/2$. On a segment, write $B$ for the selected interpolant in local time. For mesh size $h=d/M$, Duhamel's formula bounds the error from freezing $B$ at each left endpoint by $h\,\operatorname{Var}(B)$, including jumps. Replacing each frozen exponential by $I-\ii hB(jh)$ adds $O(\Lambda^2dh)$ by telescoping, since partial products have norm at most $\e^{\Lambda d}$. The strictly time-ordered discrete Dyson sum of degree $p$ is the degree-$p$ truncation of this ordered product. Its discarded terms have total norm at most
\[
\sum_{k>p}\binom{M}{k}(h\Lambda)^k
\le\sum_{k>p}\frac{(\Lambda d)^k}{k!}
\le\frac{\e^{\Lambda d}(\Lambda d)^{p+1}}{(p+1)!}.
\]
Let $S$ be the number of segments. Their variations are each at most $2\Lambda$ times the number of supplied grid intervals. Choose $M$ from this known bound and $p=O(\log(2+Sr/\eps))$ so the sum of the three local errors is at most $\xi=c\eps/(Sr)$, where $c>0$ is a sufficiently small universal constant. The fine mesh index is mapped coherently to the corresponding supplied interval, using no additional matrix queries.

For clarity, let $D$ denote a discrete Dyson segment sum and $U$ the exact grid evolution on that segment, so $\norm{D-U}\le\xi$. The sum of its absolute LCU coefficients is at most $\e^{\Lambda d}<2$. Pad its encoding to normalization $2$. One step of oblivious amplitude amplification gives the block $(3D-DD^\dagger D)/2$ at unit normalization~\cite{low2019hamiltoniansimulationinteractionpicture}. Its distance from $U$ is $O(\xi)$: replace the three factors in $DD^\dagger D$ in turn by their unitary counterparts and use $\norm{D}\le1+\xi$. Compose the segments with fresh signal ancillas. The block selected by all-zero signals is exactly the product of the amplified blocks, each a contraction. A telescoping sum gives error $O(S\xi)$ and unit normalization for this product.

Choose $c$ so the weighted simulation error is at most $\eps/(32r)$, using the uniform coefficient-weight bound. For the remaining coefficient preparation, rotations and segment gates, choose synthesis precisions so their combined operator-norm error in one LCU circuit, multiplied by $\sum_{|k|\le K}|w_k|$, is at most $\eps/(32r)$. This follows by summing individual gate errors; the supplied oracles are exact. Adding the analytic and time-grid errors gives total unnormalized operator error at most $\eps/(4r)$ for the compiled circuit. Since the target on the normalized input has norm $\norm{u_T}/\norm{u_0}\ge1/r$, Eq.~\eqref{eq:state-error} gives conditional state error at most $\eps/2$.

The approximate vector before division by the LCU weight has norm at least $(1-\eps/4)/r$. Its bounded weight therefore gives a known $\Omega(1/r^2)$ lower bound on the success probability. Use the compiled LCU and input-preparation circuit, its inverse gate sequence, and exact computational-basis reflections in the randomized amplification of Appendix~\ref{proofthm5}. This gives constant heralded success in $O(r)$ calls and preserves the successful state up to global phase. In the autonomous case a selected simulation at the same weighted precision costs $O(\alpha_AT\ell+\ell)$ queries~\cite{Low2019hamiltonian,10.1145/3313276.3316366}. In the time-dependent case the amplified Dyson circuit uses $O(p)$ queries per segment~\cite{low2019hamiltoniansimulationinteractionpicture}. Multiplying the simulation and amplification costs yields Theorem~\ref{thmhode}. The LCU index uses $O(\log(2+(\alpha_A T+\ell)\ell))$ qubits, in addition to the simulation registers.

\section{Connection to LCHS Methods}\label{lchs_conn}
This appendix uses the same $U(z)$ and $\rho_L$ as Eq.~\eqref{eq:rho}. We compare kernels at operator error $0<\delta\le 1/2$ before normalizing the output state.

\subsection{Cauchy kernel}
Set $m=1$, $p_1=\ii$ in Lemma~\ref{Matrix_Series}. The identity is
\begin{equation}\label{origin_lchs_series}
V=\frac{1-\e^{-2\pi a}}{a}
\sum_{k\in\mathbb Z}\frac{U(k/a)}{\pi(1+(k/a)^2)}
-\e^{-2\pi a}U(\ii).
\end{equation}
The last term comes from the pole at $\ii$. Let $V_K$ be the real sum in Eq.~\eqref{origin_lchs_series}, including its prefactor and restricted to $|k|\le K$. Since $\norm{U(k/a)}=1$ and $\sum_{k>K}k^{-2}\le 1/K$ for $K\ge1$, both real tails and the omitted pole give
\begin{equation}\label{origin_lchs}
\norm{V-V_K}\le\e^{\rho_L-2\pi a}+\frac{2a}{\pi K}.
\end{equation}
Choose $a=\max\{1,[\rho_L+\log(2/\delta)]/(2\pi)\}$ and $K=\lceil4a/(\pi\delta)\rceil$. Each term is at most $\delta/2$. The real coefficients are positive, and setting $H=L=0$ in Eq.~\eqref{origin_lchs_series} gives their sum as $1+\e^{-2\pi a}$. Thus $K/a=O(1/\delta)$, as for the Cauchy LCHS kernel~\cite{PhysRevLett.131.150603}. For a normalized output set $\delta=\Theta(\eps/r)$, where $r=\norm{u_0}/\norm{u_T}$. The cutoff is then $O(r/\eps)$ before the additional $O(r)$ factor for amplification.

\subsection{Sub-exponential kernel}
Fix $0<\beta<1$ and use the principal branch in
\begin{equation}
h_\beta(z)=\exp((1+\ii z)^\beta-2^\beta),\qquad
w_k^{(\beta)}=\frac{1-\e^{-2\pi a}}{2\pi a(1-\ii k/a)}
\exp(2^\beta-(1+\ii k/a)^\beta).
\end{equation}
The branch cut is $z=\ii(1+s)$, $s\ge0$. For $b=2\pi a-\rho_L\ge1$, the same residue calculation with the two sides of this cut gives
\begin{align}\label{improved_int}
V-\sum_{k\in\mathbb Z}w_k^{(\beta)}U(k/a)
={}&\frac{(\e^{-2\pi a}-1)\e^{2^\beta}}{\pi}
\int_0^\infty
\frac{U(\ii(1+s))}{(s+2)(\e^{2\pi a(1+s)}-1)}\\[-2mm]
&\hspace{10mm}\times\e^{-s^\beta\cos(\pi\beta)}
\sin(s^\beta\sin(\pi\beta))\,ds.\nonumber
\end{align}
On the right and left banks, $(1+\ii z)^\beta=s^\beta\e^{\pm\ii\pi\beta}$, respectively. The right bank runs downward, the left upward, and $dz/(z+\ii)=ds/(s+2)$. Subtracting the right-bank value of $1/h_\beta$ from the left-bank value yields $2\ii\e^{2^\beta-s^\beta\cos(\pi\beta)}\sin(s^\beta\sin(\pi\beta))$, giving Eq.~\eqref{improved_int} after division by $2\pi\ii$.

On the pole-avoiding squares of Appendix~\ref{proof01}, the lower half-plane has $\norm{U(z)}\le1$ and $\Re((1+\ii z)^\beta)\ge\cos(\pi\beta/2)|1+\ii z|^\beta$. In the upper half-plane within angle $\alpha$ of either real ray, choose $\alpha>0$ with $\beta(\pi/2+\alpha)<\pi/2$. Then $1/h_\beta$ decays as $O(\exp(-c_\beta|z|^\beta))$ for some $c_\beta>0$. On the rest of the upper contour, $\Im z\ge |z|\sin\alpha$ and the lattice denominator together with Eq.~\eqref{norm_len1} supplies $O(\e^{-b\Im z})$, dominating $\exp(|1+\ii z|^\beta)$. The outer integral vanishes, as does the small circle about $\ii$, where the integrand stays bounded.

\begin{figure}[htbp] 
		\centering 
		\begin{tikzpicture}[
			font=\small, >={Stealth[length=1mm, width=1mm]},declare function={
				R=3.5;
				r=0.05;
				offset=0.05;
			}
			]
			
			\draw[gray, ->, thin] (-R-0.5,0) -- (R+0.5,0) node[below left] {Re};
			\draw[gray,->, thin] (0,-R-0.5) -- (0,R+0.5) node[below left] {Im};
			\filldraw (0,0) circle (0.8pt) node[below right] {$O$};

			\draw[red, semithick, decorate, decoration={
				snake, segment length=5pt, amplitude=0.5pt
			}] (-r,0) -- (-R,0);

			\draw[blue, semithick, -] (-r, offset) arc[start angle=135, end angle=-135, radius={sqrt(2)*r}];
			
			\draw[blue, -, semithick] (-R, -R) -- (-R, -0.05);
			
			\draw[blue,<-, semithick] (-R, 0.7) -- (-R, R);
			\draw[blue,-, semithick] (-R, 0.7) -- (-R, 0.05);
			
			\draw[blue,-, semithick] (-R, R) -- (0.8, R);
			\draw[blue,<-, semithick] (0.7, R) -- (R, R);
			
			\draw[blue,-, semithick] (R, R) -- (R, -0.8);
			\draw[blue,<-, semithick] (R, -0.7) -- (R, -R);
			
			\draw[blue,-, semithick] (R, -R) -- (-0.8, -R);
			\draw[blue,<-, semithick] (-0.7, -R) -- (-R, -R);
			
			\draw[blue, semithick, -] (-r, offset) -- (-1, offset);
			\draw[blue, semithick, <-] (-1, offset) -- (-R, offset);

			\draw[blue, semithick, -] (-R, -offset) -- (-2, -offset);
			
			\draw[blue, semithick, <-] (-2, -offset) -- (-r, -offset);
			
			\node[black,fill=white,inner sep=1.5pt] at (0, R*0.5) {$	\oint_{\color{blue}\Gamma_{1}}{
					\frac{{(\mathrm{e}^{-2\pi a}-1)}
						\mathcal{T}\mathrm{e}^{\int_{0}^{T}(- i H(s)+(1-z)L(s))\mathrm{d}s}
					}{2\pi i (z-2)\mathrm{e}^{-2^{\beta}}\mathrm{e}^{z^{\beta}}(\mathrm{e}^{2\pi a(1-z)}-1)
					}
				}\mathrm{d}z$};
			\node[black] at (-R*0.5, 0.85) {Branch Cut};
			\node[blue] at (-R*0.85, R*0.85) {$\Gamma_1$};
			\node[black] at (-R*0.5, 0.30) {$z^{\beta}=|z|^{\beta}\mathrm{e}^{\pi\beta i }$};
			\node[black] at (-R*0.5, -0.30) {$z^{\beta}=|z|^{\beta}\mathrm{e}^{-\pi\beta i }$};
			\node[blue] at (R+0.3, R+0.3) {$+R+ i R$};
			\node[blue] at (-R-0.3, R+0.3) {$-R+ i R$};
			\node[blue] at (R+0.3, -R-0.3) {$+R- i R$};
			\node[blue] at (-R-0.3, -R-0.3) {$-R- i R$};	
		\end{tikzpicture}
		\caption{Branch-cut contour after the affine change $z_{\rm new}=1+iz_{\rm old}$. The negative real cut corresponds to the positive imaginary cut in Eq.~\eqref{improved_int}.} 
		\label{fig:tikz_rectangle_02}
	\end{figure}
Using $\e^{2\pi a(1+s)}-1\ge (1-\e^{-1})\e^{2\pi a(1+s)}$, Eq.~\eqref{norm_len1} bounds the norm of the right-hand side of Eq.~\eqref{improved_int} by
\begin{equation}
C_\beta\e^{-b}\int_0^\infty\frac{\e^{-bs+s^\beta}}{s+2}\,ds
\le C'_\beta\e^{-b},\qquad b\ge1.
\end{equation}
The integral is uniformly bounded for $b\ge1$ since $s^\beta=o(s)$. On the real axis $\Re((1+\ii x)^\beta)\ge d_\beta |x|^\beta$, with $d_\beta=\cos(\pi\beta/2)>0$. For $B=K/a\ge1$, monotone integral comparison bounds both real tails by
$C_\beta\int_B^\infty \e^{-d_\beta x^\beta}\,dx/x
\le C_\beta\e^{-d_\beta B^\beta}/(\beta d_\beta B^\beta)$.
The finite interval $|k/a|\le1$ has bounded total weight for $a\ge1$, so $\sum_k|w_k^{(\beta)}|=O_\beta(1)$.
For a sufficiently large constant $C_\beta$ depending only on $\beta$, choose
\begin{equation}
\begin{aligned}
a&=\max\{1,[\rho_L+\log(C_\beta/\delta)]/(2\pi)\},\\
K&=\left\lceil a\max\{1,[\log(C_\beta/\delta)/d_\beta]^{1/\beta}\}\right\rceil .
\end{aligned}
\end{equation}
The branch and real-tail errors then sum to at most $\delta$, with $K/a=O_\beta(\log^{1/\beta}(1/\delta))$ as in the fixed-$\beta$ LCHS bound~\cite{an2023quantumalgorithmlinearnonunitary}.

\subsection{Gaussian kernel}
Fix $c>0$ and write $g>0$ for the Gaussian width, to distinguish it from the special-function parameter $\gamma$. Set
\begin{equation}
h_2(z)=\frac{z-\ii}{-2\ii}\exp\left(\frac{z^2+1}{4g^2}-c(1-\ii z)\right),
\end{equation}
so $h_2(-\ii)=1$, and define
\begin{equation}
w_k^{(G)}=\frac{1-\e^{-2\pi a}}{a\pi(1+(k/a)^2)}
\exp\left(-\frac{(k/a)^2+1}{4g^2}+c(1-\ii k/a)\right).
\end{equation}
Apply the residue theorem on a rectangle at heights $\pm y$, $y>1$, with vertical sides sent to infinity along half-grid points. If
\begin{equation}
J(z)=\frac{(1-\e^{-2\pi a})U(z)}{\pi(z^2+1)(\e^{-2\pi\ii az}-1)}
\exp\left(-\frac{z^2+1}{4g^2}+c(1-\ii z)\right),
\end{equation}
then the exact identity is
\begin{equation}\label{Super_Eq123}
V-\sum_{k\in\mathbb Z}w_k^{(G)}U(k/a)+\e^{2c-2\pi a}U(\ii)
=\int_{-\infty}^{\infty}J(x-\ii y)\,dx
-\int_{-\infty}^{\infty}J(x+\ii y)\,dx.
\end{equation}
The residues of $2\pi\ii J$ at $-\ii$, $k/a$, and $\ii$ are $V$, $-w_k^{(G)}U(k/a)$, and $\e^{2c-2\pi a}U(\ii)$, respectively. The positively oriented upper edge runs to the left. The vertical edges vanish by Gaussian decay at fixed $y$.

	\begin{figure}[htbp]
		\centering
		\begin{tikzpicture}[
			font=\small,
			>={Stealth[length=1mm, width=1mm]}, declare function={
				R=3.5;
				r=0.05;
				offset=0.05;
			}
			]
			
			\draw[gray, ->, thin] (-R-0.5,0) -- (R+0.5,0) node[below left] {Re};
			\draw[gray,->, thin] (0,-R-0.5) -- (0,R+0.5) node[below left] {Im};
			\filldraw (0,0) circle (0.8pt) node[below right] {$O$};

			\draw[blue, -, semithick] (-R, -1.2) -- (-R, 0.8);
			\draw[blue,<-, semithick] (-R, 0.7) -- (-R, 1.2);
			
			\draw[blue,-, semithick] (-R, 1.2) -- (0.8, 1.2);
			\draw[blue,<-, semithick] (0.7, 1.2) -- (R, 1.2);
			
			\draw[blue,-, semithick] (R, 1.2) -- (R, -0.8);
			\draw[blue,<-, semithick] (R, -0.7) -- (R, -1.2);
			
			\draw[blue,-, semithick] (R, -1.2) -- (-0.8, -1.2);
			\draw[blue,<-, semithick] (-0.7, -1.2) -- (-R, -1.2);

			\node[blue,anchor=west,xshift=2pt] at (0, 1.36) {$ i y$};
			\node[blue,anchor=west,xshift=2pt] at (0, -1.36) {$- i y$};
			\node[blue] at (-R*0.85, R*0.25) {$\Gamma_2$};
			\node[black,fill=white,inner sep=1.5pt] at (0, R*0.65) {$\int_{\color{blue}\Gamma_{2}}{
					\frac{({\mathrm{e}^{-2\pi a}-1})\mathcal{T}\mathrm{e}^{- i \int_{0}^{T}(H(s)+zL(s))\mathrm{d}s}}{2\pi i (z+ i )(\mathrm{e}^{-2\pi  i az}-1)h_{2}(z)}\mathrm{d}z,
				}$};
			\node[blue] at (R+0.3, 1.2+0.3) {$+R+ i y$};
			\node[blue] at (-R-0.3, 1.2+0.3) {$-R+ i y$};
			\node[blue] at (R+0.3, -1.2-0.3) {$+R- i y$};
			\node[blue] at (-R-0.3, -1.2-0.3) {$-R- i y$};	
		\end{tikzpicture}
		\caption{Schematic diagram of the matrix contour integral in Eq.~\eqref{Super_Eq123}.}
		\label{fig:tikz_rectangle_03}
	\end{figure}
Take $a\ge1$, $y=2cg^2\ge2$, and $b=2\pi a-\rho_L\ge2c$. On either horizontal edge,
$|z^2+1|=|z-\ii||z+\ii|\ge x^2+(y-1)^2$, whose reciprocal integrates to at most $\pi$ for $y\ge2$. The bottom-edge lattice ratio is at most $(1-\e^{-2\pi a})/(1-\e^{-2\pi ay})\le1$; the top-edge ratio has the extra factor $\e^{-2\pi ay}$. Using Eq.~\eqref{norm_len1} gives
\begin{align}
\int_{\mathbb R}\norm{J(x-\ii y)}\,dx&\le\e^{c-c^2g^2},\\
\int_{\mathbb R}\norm{J(x+\ii y)}\,dx&\le\e^{c+3c^2g^2-2bcg^2}
\le\e^{c-c^2g^2},\\
\norm{\e^{2c-2\pi a}U(\ii)}&\le\e^{2c-b}.
\end{align}
By Eq.~\eqref{origin_lchs_series} with $H=L=0$, the discrete weight obeys
\begin{equation}
\sum_k|w_k^{(G)}|\le\frac{(1-\e^{-2\pi a})\e^c}{a\pi}\sum_k\frac1{1+(k/a)^2}
=\e^c(1+\e^{-2\pi a}).
\end{equation}
Retaining the Gaussian factor for $|k|>K$ bounds the real tail by $2\e^c\exp[-(K/a)^2/(4g^2)]$. One sufficient choice is
\[
\begin{aligned}
g^2&=\max\{1/c,[c+\log(8/\delta)]/c^2\},\\
a&=\max\{1,[\rho_L+2c+\log(8/\delta)]/(2\pi)\},\\
K&=\left\lceil 2ag\sqrt{\log(8\e^c/\delta)}\right\rceil .
\end{aligned}
\]
Each horizontal-edge error and the auxiliary-pole error are at most $\delta/8$, and the real tail is at most $\delta/4$. For fixed $c$, $K/a=O_c(\log(1/\delta))$ and the total weight is constant, agreeing in leading scaling with the Gaussian LCHS construction~\cite{low2025optimalquantumsimulationlinear}.

\section{Proof of Non-Hermitian Wave Equation Theorem~\ref{wavethm}}\label{proofthm5}
This appendix proves Theorem~\ref{wavethm} and the estimates used in Theorems~\ref{besselthm} and \ref{airythm}. All matrix-query counts in these three proofs refer to the two supplied block-encodings of $H/\norm H$ and $L/\norm L$. The common class is
\begin{equation}\label{eq:cos-mixture}
G(X)=\int_0^1\cos(\tau s\sqrt X)\,d\mu(s),\qquad
\mu\text{ a probability measure},\quad \tau>0.
\end{equation}
The even cosine series converges absolutely and uniformly on bounded matrix sets, uniformly in $s$, and defines an entire function of $X$. For Airy, $G_t(X)=f(tX^{1/3})/f(0)$, with the unnormalized $f$ retained from the main text. The wave propagator has a unit point mass at $s=1$. For $X\succeq0$, the spectral theorem gives $\norm{G(X)}\le1$.

\subsection{Parameters used only in the proof}
For the query bounds set $\gamma=\norm H$ and let $\tau=t$ for wave and Bessel, or $\tau=2t^{3/2}/3$ for Airy. Put $x_k=(k+1/2)\pi$, $s_k=x_k^2/(\gamma\tau^2)$, and $M_s=L+sH$ for $s\ge0$. The sampled matrices are $M_{s_k}$. The regulator and coefficients are $h(z)=\cos(\tau\sqrt{i\gamma z})$ and $d_k=2(-1)^k/[x_k(1+is_k)]$, agreeing with the main-text formulas.
For $|\psi_0\rangle=u_0/\norm{u_0}$, assume a known $\eta>0$ such that $\eta\le\norm{G(A)|\psi_0\rangle}=\norm{u(t)}/\norm{u_0}\le C\eta$ for a fixed universal constant $C$. For Airy, the definition in Eq.~\eqref{eq:airy-ur} gives $\exp(\tau\sqrt{\gamma/2})/\norm{G_t(A)|\psi_0\rangle}=|f(0)|u_r$. For wave and Bessel, the same ratio equals $u_r$. Since $f(0)=\Ai(0)>0$ is a fixed universal constant, it is absorbed into the bounds below. If no sampling terms are retained, use $K=-1$ and an empty sum. Otherwise choose $K\ge0$ to satisfy
\begin{equation}\label{eq:Kchoice}
\frac{|h(1)|\gamma\tau^2}{\pi^3(K+1/2)^2}\le\frac{\eps\eta}{8}.
\end{equation}
With $b(s)=\norm L+s\norm H$ for $s\ge0$, define
\begin{align}\label{eq:special-parameters}
S_K&=|h(1)|\left(1+\sum_{k=0}^K|d_k|\right),\\
b_K&=\begin{cases}b(0),&K=-1,\\b(s_K),&K\ge0.\end{cases}\\
D_K&=O\left(\kappa\left[\tau\sqrt{b_K}
+\log\left(2+\frac{\kappa S_K}{\eps\eta}\right)\right]\right).\label{eq:Dchoice}
\end{align}
The normalization of the implemented finite sum is $4S_K$. The cutoff choice, including the small-time empty-sum case, is justified below.

\subsection{A contour bound for noncommuting matrices}
Let $H\succ0$ and $L\succ0$ be fixed. For $|w|$ sufficiently large, the principal square root
\begin{equation}
B_w=(H+w^{-1}L)^{1/2}
\end{equation}
is analytic in $w^{-1}$ near zero. Choose a fixed contour in the open right half-plane enclosing $\sigma(H)$. For sufficiently large $|w|$, the Neumann series bounds the resolvent of $H+L/w$ uniformly on this contour. The resolvent identity in the Cauchy integral for the square root then gives
\begin{equation}\label{eq:root-expand}
B_w=H^{1/2}+E_w,\qquad \norm{E_w}\le C/|w|.
\end{equation}
This is an operator estimate and does not assume $[H,L]=0$. For any square-root choice of the scalar $w$, $\sqrt w B_w$ squares to $L+wH$. Its use inside the even cosine series therefore represents the entire function in Eq.~\eqref{eq:cos-mixture}.

For any matrix $B$, differentiating $\norm{\exp(\pm\ii v B)y}^2$ with respect to $v\ge0$ and using the Hermitian part of $\pm\ii B$ gives, at $v=1$,
\begin{equation}
\norm{\exp(\pm\ii B)}\le\exp\left(\norm{(B-B^\dagger)/(2\ii)}\right).
\end{equation}
Using Eq.~\eqref{eq:root-expand}, for $|w|=R$ and $0\le s\le1$,
\begin{equation}
\norm{\cos(\tau s\sqrt w B_w)}
\le\exp\left(\tau\sqrt{\norm H}|\Im\sqrt w|+C\tau/\sqrt R\right).
\end{equation}
Integrating against the probability measure gives the same bound for $G(L+wH)$.

The contour argument holds for every $\gamma\ge\norm H$. Take $R_n=(n\pi/(\tau\sqrt\gamma))^2$ for positive integers $n$ sufficiently large that $R_n>2$ and Eq.~\eqref{eq:root-expand} applies. For $|w|=R_n$, write $q=\tau\sqrt{\gamma w}$, so $|q|=n\pi$. There is a constant $c_0>0$, independent of $n$, such that
\begin{equation}\label{eq:cos-denom}
|\cos q|\ge c_0\exp(|\Im q|).
\end{equation}
Indeed, if $|\Im q|\ge1$, use $|\cos q|^2=\cos^2(\Re q)+\sinh^2(\Im q)$. If $|\Im q|\le1$, then $0\le n\pi-|\Re q|=(\Im q)^2/(n\pi+|\Re q|)\le1/(n\pi)$, hence $|\cos(\Re q)|\ge\cos(1/\pi)$.

Now set $w=\ii z$ and $h(z)=\cos(\tau\sqrt{\ii\gamma z})$. Equations~\eqref{eq:root-expand}--\eqref{eq:cos-denom} imply
\begin{equation}
\sup_{|z|=R_n}\frac{\norm{G(L+\ii zH)}}{|h(z)|}=O(1).
\end{equation}
Consequently $F(z)=h(1)G(L+\ii zH)/[h(z)z(z-1)]$ has norm $O(R_n^{-2})$ on these circles, and its contour integral is $O(R_n^{-1})\to0$. The same estimate holds at $\gamma=\norm H$. Its implicit constant may depend on the fixed matrices and $\tau$. The explicit residue tail below controls the implementation error independently of this constant.

\subsection{Residues and the coefficient tail}
The residues at $z=1$ and $z=0$ are $G(A)$ and $-h(1)G(L)$. At each simple zero $z_k=-\ii s_k$ of $h$, differentiation gives $h'(z_k)=-\ii\gamma\tau^2(-1)^k/(2x_k)$ and
\begin{equation}
h'(z_k)z_k(z_k-1)=(-1)^k\frac{x_k}{2}(1+\ii s_k),
\end{equation}
so the residue is $h(1)d_kG(M_{s_k})$ with $d_k$ from Eq.~\eqref{eq:regulator}. The circle of radius $R_n$ encloses $0,1$ and exactly the zeros $z_0,\ldots,z_{n-1}$ of $h$. A removable singularity of $F$ contributes zero. Taking $n\to\infty$ in its residue formula gives Eq.~\eqref{eq:lcu_exact}. The sample series is absolutely convergent by the bound below.

Since $M_{s_k}\succ0$ and $\norm{G(M_{s_k})}\le1$, for $K\ge0$,
\begin{equation}\label{eq:tail-bound}
\left\|h(1)\sum_{k>K}d_kG(M_{s_k})\right\|
\le |h(1)|\frac{2\gamma\tau^2}{\pi^3}
\sum_{k>K}\frac1{(k+1/2)^3}
\le\frac{|h(1)|\gamma\tau^2}{\pi^3(K+1/2)^2}.
\end{equation}
The last inequality follows by comparison with the integral starting at $K+1/2$. All prefactors in norm estimates use $|h(1)|$, since $h(1)$ is generally complex.

The two coefficient bounds
\begin{equation}
|d_k|\le\min\left\{\frac2{\pi(k+1/2)},
\frac{2\gamma\tau^2}{\pi^3(k+1/2)^3}\right\}
\end{equation}
give, by splitting near $k+1/2\simeq\tau\sqrt\gamma/\pi$,
\begin{equation}\label{eq:estmdk}
1+\sum_{k=0}^\infty|d_k|=O\bigl(\log(2+\gamma\tau^2)\bigr).
\end{equation}
When $\tau\sqrt\gamma/\pi<1/2$, the cubic bound controls the whole sum. Also
\begin{equation}\label{eq:hmodulus}
|h(1)|=\left|\cos\left(\frac{1+\ii}{\sqrt2}\tau\sqrt\gamma\right)\right|
\le\exp(\tau\sqrt{\gamma/2}),
\end{equation}
and it is asymptotic to one half of this exponential for large $\tau\sqrt\gamma$. Equations~\eqref{eq:tail-bound}--\eqref{eq:hmodulus} prove the truncation and normalization estimates in the main text.

\subsection{A globally bounded polynomial approximation}
The separate-oracle LCU in Eq.~\eqref{eq:separate-oracles} encodes $M_s=L+sH$ with normalization $b(s)=\norm L+s\norm H$, for $s\ge0$. To see the circuit explicitly, prepare amplitudes $\sqrt{\norm L/b(s)}$ and $\sqrt{s\norm H/b(s)}$, select the normalized $L$ and $H$ oracles, and unprepare the same real state. All weights are nonnegative. Thus
\begin{equation}\label{eq:gap}
\frac{\lambda_{\min}(M_s)}{b(s)}
\ge\frac{\lambda_{\min}(L)+s\lambda_{\min}(H)}{\norm L+s\norm H}
\ge\frac1{\max(\kappa_L,\kappa_H)}=\frac1\kappa,
\end{equation}
and the upper endpoint is at most one. Equation~\eqref{eq:gap} uses the separate normalization of the two supplied oracles.

In choosing the approximation interval, use the computable upper bound on $\kappa$ from Eq.~\eqref{eq:kappabe}. It changes the following estimates only by a constant factor under the stated spectral promise. With this convention, let
$g=\min\{1/2,\kappa^{-1}\}$,
$x_0=(1+g)/2$, $r_0=(1-g)/2$, and buffer $\Delta=g/4$. For each retained $s$ (including $s=0$), $b(s)\le b_K$ and the function $F_s(x)=G(b(s) x)$ is entire and satisfies
\begin{equation}
|F_s(z)|\le\exp(\tau\sqrt{b(s)|z|}).
\end{equation}
On the circle $|z-x_0|=1/2$ the right-hand side is at most $\exp(C\tau\sqrt{b_K})$. Cauchy's coefficient estimate for
$F_s(x_0+u)=\sum_n a_nu^n$ therefore gives
\begin{equation}
\sum_n|a_n|(r_0+\Delta)^n
\le\frac{2}{g}\exp(C\tau\sqrt{b_K})=:B.
\end{equation}
Apply the local Taylor approximation result for bounded polynomials~\cite[Corollary 66]{10.1145/3313276.3316366} with accuracy $\widehat\delta_q=\min\{\delta_q/2,1/(2B)\}$. It yields a real polynomial $P_s$ with error at most $\delta_q/2$ on $[g,1]$, degree
\begin{equation}\label{eq:poly-degree}
D=O\left(\kappa
\left[\tau\sqrt{b_K}+\log\frac{2\kappa}{\delta_q}\right]\right),
\end{equation}
and $\sup_{[-1,1]}|P_s|\le1+\delta_q/2$. The latter bound uses $|F_s(x)|\le1$ on the enlarged \emph{positive} real interval $[g-\Delta/2,1+\Delta/2]$. Using $\widehat\delta_q$ only changes the constant multiplying $\log B$ in the degree estimate. Taking the real part of the resulting polynomial preserves all the required bounds.

For $\delta_q\le1/2$, $P_s/4$ is bounded by $1/2$ on $[-1,1]$. Apply the arbitrary-parity Hermitian polynomial transformation~\cite[Theorem 56]{10.1145/3313276.3316366} with circuit approximation error at most $\delta_q/8$. With ideal state-preparation rotations, Eq.~\eqref{eq:separate-oracles} is an exact input block-encoding. Finite synthesis of these rotations is accounted for in the complete LCU circuit below. If $\widetilde P_s$ is the resulting top block, then
\begin{equation}\label{eq:qsvt-unscaled-error}
\norm{4\widetilde P_s-G(M_s)}
\le 4\norm{\widetilde P_s-P_s(M_s/b(s))/4}
+\norm{P_s(M_s/b(s))-G(M_s)}
\le\delta_q.
\end{equation}
Thus this is a normalization-four block-encoding of $G(M_s)$ with unnormalized error $\delta_q$, using $O(D)$ input queries. The polynomial approximation contributes at most $\delta_q/2$, and circuit synthesis contributes at most $4(\delta_q/8)$ in the same unnormalized convention. 

\subsection{LCU errors, output normalization, and queries}
Choose $K$ by the cutoff rule below, which makes the tail at most $\eps\eta/8$. Each exact sample has norm at most one, so the retained sum has norm at most $S_K$. Hence $S_K\ge\norm{G(A)|\psi_0\rangle}-\eps\eta/8\ge(1-\eps/8)\eta$. Set
\begin{equation}\label{eq:qsvt-budget}
\delta_q=\frac{\eps\eta}{16S_K}<\frac12.
\end{equation}
The weighted function error in Eq.~\eqref{eq:qsvt-unscaled-error} is at most $S_K\delta_q=\eps\eta/16$. Compile the remaining gates of the complete LCU circuit, including every use of the sample-dependent input rotations, their inverses and the outer coefficient preparations, to total operator-norm error at most $\eps\eta/(64S_K)$. This follows by making the sum of the individual gate errors at most this tolerance. Multiplying the top block by $4S_K$ then contributes at most $\eps\eta/16$. The resulting target approximation satisfies
\begin{equation}
\norm{\widetilde G-G(A)}\le\eps\eta/4,
\end{equation}
and is encoded with normalization $4S_K$. For $v=G(A)|\psi_0\rangle$ and $\widetilde v=\widetilde G|\psi_0\rangle$, the elementary normalization estimate is
\begin{equation}\label{eq:state-error}
\left\|\frac{v}{\norm v}-\frac{\widetilde v}{\norm{\widetilde v}}\right\|
\le\frac{2\norm{v-\widetilde v}}{\norm v}\le\eps/2.
\end{equation}
Let $W$ be the compiled LCU circuit composed with the clean input-state preparation. Project every auxiliary and work register onto its all-zero state. The successful component of $W|0\cdots0\rangle$ is then $|0\cdots0\rangle\otimes\widetilde v/(4S_K)$ and its probability $p$ satisfies $p\ge(\eta/(8S_K))^2$. Use $W$ and its exact inverse gate sequence for amplitude amplification. Both the all-input-zero reflection and the success reflection are computational-basis phase flips, implementable exactly using $X$, Toffoli, and $Z$ gates. Thus each successful run yields the pure system state $\widetilde v/\norm{\widetilde v}$ up to global phase.

For completeness, the supplied lower bound on $p$ suffices to choose the number of queries in randomized amplitude amplification~\cite{Brassard2002Amplitude}. With probability $1/2$ use $W$ alone. Otherwise choose $j$ uniformly from $0,\ldots,M-1$, with $M=\lceil8S_K/\eta\rceil$, and perform $j$ Grover iterations. For $p=\sin^2\theta$ the resulting success probability is $\sin^2((2j+1)\theta)$. If $p\ge1/2$, the first choice succeeds with overall probability at least $1/4$. For $p<1/2$ the mean success probability over $j$ is $1/2-\sin(4M\theta)/(4M\sin(2\theta))$, and $M\sin(2\theta)\ge\sqrt2$. This gives constant overall success using $O(S_K/\eta)$ calls to the circuit and input preparation. Every successful choice of $j$ gives the same system-state direction, so Eq.~\eqref{eq:state-error} applies after heralding.

To select the transformations coherently, pad the work registers to a common size and control each query and phase on the sampling index. At each of the $O(D_K)$ steps a constant number of controlled forward and inverse oracle calls covers all active branches. Phase selection and coefficient arithmetic enter the gate cost. Equation~\eqref{eq:qsvt-budget} in Eq.~\eqref{eq:poly-degree} gives Eq.~\eqref{eq:Dchoice}.

These bounds use exact supplied oracles. For any fixed complete amplified unitary circuit, $Q$ calls with unitary operator-norm error at most $\zeta$ introduce at most $Q\zeta$ unitary error, counting all controlled and inverse calls. A bound on the encoded matrix block alone requires the corresponding QSVT robustness estimate.

\subsection{Eliminating the cutoff and polynomial degree}
For an empty sample sum, the full coefficient tail obeys
\begin{equation}\label{eq:empty-tail}
\left\|h(1)\sum_{k\ge0}d_kG(M_{s_k})\right\|
\le\frac{14\zeta(3)}{\pi^3}|h(1)|\gamma\tau^2
<|h(1)|\gamma\tau^2.
\end{equation}
Here $\sum_{k\ge0}(k+1/2)^{-3}=7\zeta(3)$. If
$\gamma\tau^2\exp(\tau\sqrt{\gamma/2})\le\eps\eta/8$, set $K=-1$ and retain $h(1)G(L)$. Otherwise choose
\begin{equation}
K=\left\lceil\sqrt{\frac{8\gamma\tau^2\exp(\tau\sqrt{\gamma/2})}{\pi^3\eps\eta}}\right\rceil.
\end{equation}
This ensures Eq.~\eqref{eq:Kchoice} since $|h(1)|\le\exp(\tau\sqrt{\gamma/2})$. The quantity under the square root exceeds $1/\pi^3$ in this branch, so increasing the root by at most $3/2$ to obtain $K+1/2$ changes it by at most a constant factor. As $\eta$ is a constant-factor lower bound on the normalized-input output norm, $\exp(\tau\sqrt{\gamma/2})/\eta=\Theta(u_r)$ for all three propagators, with the fixed Airy factor given above. Thus $K+1/2=O(\tau\sqrt{\gamma u_r/\eps})$ when $K\ge0$. Here $b_K-\norm L=\pi^2(K+1/2)^2\norm H/(\gamma\tau^2)$, whereas $b_{-1}=\norm L$. In both branches,
\begin{equation}
b_K=O\bigl(\norm L+\norm H\,u_r/\eps\bigr),\qquad
\frac{S_K}{\eta}=O\bigl(u_r\log(2+\gamma\tau^2)\bigr).
\end{equation}
Substituting into the polynomial bound gives the explicit estimate
\begin{align}\label{eq:expanded-cost-proof}
Q_H+Q_L=O\Biggl(&\kappa u_r\log(2+\gamma\tau^2)\Biggl[
\tau\sqrt{\norm L+\norm H\,u_r/\eps}\\
&\hspace{12mm}+\log\left(2+\frac{\kappa u_r\log(2+\gamma\tau^2)}{\eps}\right)\Biggr]\Biggr).
\end{align}
Together with the input-state count $O(S_K/\eta)$, this gives the three theorem statements. At $t=0$ each propagator is $I$, and the input state may be returned directly.

\subsection{Cost of the direct Taylor--LCU baseline}
For wave, Bessel $J_0$, and Airy dynamics, the direct Taylor circuit builds its block of $A/(\norm L+\norm H)$ from the same two supplied blocks, $H/\norm H$ and $L/\norm L$, using a two-term LCU. This costs $O(1)$ calls to the supplied oracles. The full scalar absolute-coefficient sums are
\begin{equation}\label{eq:taylor-sums}
\cosh(t\sqrt{\norm L+\norm H}),\quad
I_0(t\sqrt{\norm L+\norm H}),\quad
{}_0F_1\left(;\frac23;\frac{(\norm L+\norm H)t^3}{9}\right)
\end{equation}
for wave, Bessel and normalized Airy respectively. In the variable $x=\tau\sqrt{\norm L+\norm H}$, the Bessel and Airy coefficient ratios to the coefficient $x^{2n}/(2n)!$ of $\cosh x$ are $(1/2)_n/(1)_n\le1$ and $(1/2)_n/(2/3)_n\le1$. For $2n\ge e^2x$, Stirling's bound gives $x^{2n}/(2n)!\le e^{-2n}$. Consequently a monomial degree
\begin{equation}
O\left(1+\tau\sqrt{\norm L+\norm H}+\log\left(2+\frac1{\eps\eta}\right)\right)
\end{equation}
suffices to make the omitted absolute-coefficient sum, and hence the operator error, at most $\eps\eta/8$ by the geometric tail bound. Each full sum in Eq.~\eqref{eq:taylor-sums} is at least $\norm{G(A)|\psi_0\rangle}\ge\eta$. The retained sum is therefore at least $1-\eps/8$ times the full sum. A single unsegmented LCU of these monomials uses the retained sum as its normalization. Its successful output norm before division by this normalization differs from $\norm{G(A)|\psi_0\rangle}$ by at most $\eps\eta/8$. Thus the retained sum divided by $\eta$ bounds the number of amplitude-amplification calls up to a constant. Each call needs $O(1)$ input-state queries and a number of matrix queries at most proportional to the monomial degree. The wave matrix-query upper bound is
\begin{align}\label{eq:taylor-wave-queries}
O\Biggl(&\frac{\norm{u_0}}{\norm{u(t)}}\cosh(t\sqrt{\norm L+\norm H})\\
&\times\left[1+t\sqrt{\norm L+\norm H}+\log\left(2+\frac{\norm{u_0}}{\eps\norm{u(t)}}\right)\right]\Biggr).
\end{align}
The comparison concerns this monomial construction with the stated absolute-tail budget. It gives no lower bound on other LCU constructions or on general algorithms.

For an explicit comparison, take $t=1$, $H=I$, $L=\operatorname{diag}((2\pi n)^2,((2n+1)\pi)^2)$, and $|u_0\rangle=(1,1)^T/\sqrt2$. Then $\kappa=O(1)$, $\norm L=\Theta(n^2)$ and $\norm{u(1)}/\norm{u_0}\to1$. Indeed, $\sqrt{(m\pi)^2+\ii}=m\pi+O(1/m)$, so the wave components approach $1/\sqrt2$ and $-1/\sqrt2$, respectively. At fixed precision the CBMD matrix-query upper bound is $\widetilde O(n)$. For the stated Taylor construction the retained sum grows as $\exp(\Theta(n))$, so the unamplified success probability is $\exp(-\Theta(n))$ even though the output norm tends to one. For ordinary amplitude amplification a run with $j$ iterations has success amplitude at most $(2j+1)\sqrt p$ for initial success probability $p$. Thus raising the success probability of this particular Taylor circuit to a constant costs exponentially many calls in $n$. The example also admits shortcuts if its diagonal entries are used directly. At fixed matrices, time and nonzero output, the direct Taylor degree bound grows logarithmically in $1/\eps$, whereas the present CBMD bound retains an $\eps^{-1/2}$ term.

\section{Proof of Non-Hermitian Bessel Equation Theorem~\ref{besselthm}}\label{besselproof}
The power series in Eq.~\eqref{eq:bessel-propagator} converges for every matrix. If its $t^{2n}$ coefficient is $C_n$, then
$4n^2C_n=-AC_{n-1}$ for $n\ge1$. Substitution into Eq.~\eqref{eq:bessel} proves the ODE for $t>0$, with $G_0=I$ and $G'_0=0$. For uniqueness at the singular endpoint, take the difference $w$ of two solutions that are $C^1$ at zero and $C^2$ for $t>0$. Integrating $(tw')'=-tAw$ twice with $w(0)=0$ and $\lim_{t\downarrow0}tw'(t)=0$ gives
\[
w(t)=-\int_0^t s\log(t/s)A w(s)\,ds.
\]
Since $\int_0^t s\log(t/s)\,ds=t^2/4$, the supremum of $\norm w$ on $[0,T]$ is at most $T^2\norm A/4$ times itself. Thus $w=0$ for small $T>0$, and ordinary ODE uniqueness extends the conclusion to all $t>0$.

Equation~\eqref{eq:bessel-cos} follows either from the Bessel integral~\cite{NISTDLMF} or by integrating the absolutely convergent cosine series termwise. Because cosine is even, the pushforward of $d\theta/\pi$ on $[0,\pi]$ under $\theta\mapsto|\cos\theta|$ has density $2/(\pi\sqrt{1-s^2})$ on $(0,1)$. Its $2n$th moment is $(1/2)_n/(1)_n$, which gives Eq.~\eqref{eq:bessel-propagator} term by term. Thus the common form Eq.~\eqref{eq:cos-mixture} holds with $\tau=t$.

Every step of Appendix~\ref{proofthm5} now applies without changing the regulator or the coefficients: the large-contour bound is an operator estimate; the sample matrices are positive definite; the residue sum carries a minus sign; the tail is bounded by Eq.~\eqref{eq:tail-bound}; and the QSVT errors are weighted by $S_K$, not only by $|h(1)|$. This proves Theorem~\ref{besselthm}. The naive monomial coefficient sum is
\begin{equation}
\sum_{n=0}^\infty\frac{((\norm L+\norm H)t^2/4)^n}{(n!)^2}=I_0(t\sqrt{\norm L+\norm H}).
\end{equation}
For the boundary example $A=\ii h$ with fixed $h>0$, take $y=t\sqrt{\ii h}$ on the ray $\arg y=\pi/4$. There $|J_0(y)|\sim e^{\Im y}/\sqrt{2\pi|y|}$ as $t\to\infty$~\cite[Sec.~10.17]{NISTDLMF}. Since $\Im y=t\sqrt{h/2}$, the envelope ratio grows as $\Theta(t^{1/2})$.

\section{Proof of Non-Hermitian Airy Dynamics Theorem~\ref{airythm}}\label{proof_airy}
The Airy equation gives $(d/dz)^2\Ai(-z)=-z\Ai(-z)$. In its Taylor series the coefficients of $z^{3n+2}$ vanish. The root-of-unity average in Eq.~\eqref{eq:airy-projection} retains exactly the coefficients of $z^{3n}$, since $\sum_{j=0}^2\omega^{jm}/3$ is one for $3\mid m$ and zero otherwise. Writing $c_n$ for the coefficient of $z^{3n}$ gives $(3n)(3n-1)c_n=-c_{n-1}$ and $c_0=\Ai(0)$. Thus
\begin{equation}\label{eq:airy-series-proof}
\frac{f(z)}{f(0)}=\sum_{n=0}^\infty\frac{(-z^3/9)^n}{n!(2/3)_n},
\quad f(0)=\Ai(0),\quad f'(0)=0.
\end{equation}
Stirling's formula gives $\log(n!(2/3)_n)=2n\log n+O(n)$. The coefficient formula for the order of an entire function thus gives order $3/2$ in $z$ and $1/2$ in $z^3$. Substitute $z^3=t^3A$ in this entire series. For the resulting propagator $G_t(A)=f(A^{1/3}t)/f(0)=\sum_{n\ge0}C_nt^{3n}$ the coefficient identity is
\begin{equation}
(3n)(3n-1)C_n=-AC_{n-1},\quad C_0=I.
\end{equation}
Termwise differentiation gives $G''_t(A)+tAG_t(A)=0$, $G_0(A)=I$, and $G'_0(A)=0$. The inverse output-norm factor for a unit input is $\norm{u_0}/\norm{G_t(A)u_0}$. Multiplying it by the Airy exponential in Theorem~\ref{airythm} gives exactly $|f(0)|u_r$, since $\norm{f(A^{1/3}t)u_0}=|f(0)|\norm{G_t(A)u_0}$.

The identity
\begin{equation}
{}_0F_1(;2/3;-y^2/4)
=\Gamma(2/3)(y/2)^{1/3}J_{-1/3}(y)
\end{equation}
is understood through its entire even series. Poisson's integral for $J_{-1/3}$, whose order obeys $-1/3>-1/2$, gives a probability cosine mixture with $y=(2/3)t^{3/2}\sqrt X$~\cite[Sec.~10.9]{NISTDLMF}. Its normalization follows from
\begin{equation}
\int_0^1(1-s^2)^{-5/6}ds
=\tfrac12B(1/2,1/6)
=\frac{\sqrt\pi\Gamma(1/6)}{2\Gamma(2/3)}.
\end{equation}
With $u=s^2$, the normalized $2n$th moment is
\[
\frac{B(n+1/2,1/6)}{B(1/2,1/6)}=\frac{(1/2)_n}{(2/3)_n}.
\]
Termwise integration of the cosine series therefore gives
\[
\frac{(-y^2)^n}{(2n)!}\,\frac{(1/2)_n}{(2/3)_n}
=\frac{(-y^2/4)^n}{n!(2/3)_n},
\]
where $(2n)!=4^n n!(1/2)_n$. Substituting $y^2/4=t^3X/9$ reproduces Eq.~\eqref{eq:airy-series-proof}. Uniform convergence on compact sets gives the matrix formula
\begin{equation}
G_t(X)=\frac{2\Gamma(2/3)}{\sqrt\pi\Gamma(1/6)}
\int_0^1\frac{\cos\bigl((2/3)t^{3/2}s\sqrt X\bigr)}{(1-s^2)^{5/6}}\,ds.
\end{equation}

Apply Appendix~\ref{proofthm5} with $\tau=(2/3)t^{3/2}$. The poles of $1/h$ and the derivative of $h$ satisfy
\begin{equation}
z_k=-\ii\frac{9(k+1/2)^2\pi^2}{4\gamma t^3},\qquad
h'(z_k)z_k(z_k-1)=(-1)^k\frac{(k+1/2)\pi}{2}(1+\ii s_k).
\end{equation}
The origin residue is $-h(1)G_t(L)$, the target residue is $G_t(A)$, and the sample residues are $h(1)d_kG_t(M_{s_k})$. The contour integral vanishes along the specified circles, so
\begin{equation}
G_t(A)=h(1)G_t(L)-h(1)\sum_{k=0}^K d_kG_t(M_{s_k})
-h(1)\sum_{k>K}d_kG_t(M_{s_k}).
\end{equation}
Thus the finite sum and the tail both have minus signs.
For $K\ge0$, the explicit tail and modulus estimates specialize to
\begin{align}
\norm{\text{tail}}&\le\frac{4|h(1)|\gamma t^3}{9\pi^3(K+1/2)^2},\\
|h(1)|&\le\exp\left(\frac23t^{3/2}\sqrt{\gamma/2}\right),\\
S_K&=O\left(\exp\left(\frac23t^{3/2}\sqrt{\gamma/2}\right)
\log(2+\gamma t^3)\right).
\end{align}
The actual input normalization of the largest sample is
\begin{equation}
b_K=\norm L+\frac{9(K+1/2)^2\pi^2\norm H}{4\gamma t^3},\qquad K\ge0.
\end{equation}
The cutoff rule of Appendix~\ref{proofthm5} gives $K+1/2=O(t^{3/2}\sqrt{\gamma u_r/\eps})$ whenever a sample is retained. Hence $b_K=O(\norm L+\norm H\,u_r/\eps)$, also valid with $b_{-1}=\norm L$ for the empty sum. Substitution into Eq.~\eqref{eq:Dchoice}, followed by $\tau=(2/3)t^{3/2}$ and $\gamma=\norm H$ in Eq.~\eqref{eq:expanded-cost-proof}, gives Theorem~\ref{airythm}.

Finally, the absolute sum of the coefficients of the direct Taylor construction is
\begin{equation}
{}_0F_1\left(;\frac23;\frac{(\norm L+\norm H)t^3}{9}\right).
\end{equation}
For the boundary example $A=\ii h$ with fixed $h>0$, take $y=(2/3)t^{3/2}\sqrt{\ii h}$. On this ray $|J_{-1/3}(y)|\sim e^{\Im y}/\sqrt{2\pi|y|}$~\cite[Sec.~10.17]{NISTDLMF}. The additional factor $|(y/2)^{1/3}|$ in the Airy propagator gives a prefactor proportional to $|y|^{-1/6}$, hence to $t^{-1/4}$. The resulting exponential envelope divided by $|G_t(\ii h)|$ grows as $\Theta(t^{1/4})$.

\end{document}